\documentstyle[12pt]{article}

\begin{document}

\title{A General Definition of ``Conserved Quantities'' in General 
Relativity and Other Theories of Gravity}
\author{Robert M. Wald and Andreas Zoupas\\
         {\it Enrico Fermi Institute and Department of Physics}\\
         {\it University of Chicago}\\
         {\it 5640 S. Ellis Avenue}\\
         {\it Chicago, Illinois 60637-1433}}
\maketitle

\begin{abstract}

In general relativity, the notion of mass and other conserved
quantities at spatial infinity can be defined in a natural way via the
Hamiltonian framework: Each conserved quantity is associated with an
asymptotic symmetry and the value of the conserved quantity is defined
to be the value of the Hamiltonian which generates the canonical
transformation on phase space corresponding to this symmetry. However,
such an approach cannot be employed to define ``conserved quantities''
in a situation where symplectic current can be radiated away (such as
occurs at null infinity in general relativity) because there does not,
in general, exist a Hamiltonian which generates the given asymptotic
symmetry. (This fact is closely related to the fact that the desired
``conserved quantities'' are not, in general, conserved!) In this
paper we give a prescription for defining ``conserved quantities''
by proposing a modification of the equation that must be
satisfied by a Hamiltonian. Our prescription is a very general one,
and is applicable to a very general class of asymptotic conditions in
arbitrary diffeomorphism covariant theories of gravity derivable from
a Lagrangian, although we have not investigated existence and
uniqueness issues in the most general contexts. In the case of general
relativity with the standard asymptotic conditions at null infinity,
our prescription agrees with the one proposed by Dray and Streubel
from entirely different considerations.

\end{abstract}
\newpage

\section{Introduction}

Notions of energy and angular momentum have played a key role in
analyzing the behavior of physical theories. For theories of fields in
a fixed, background spacetime, a locally conserved
stress-energy-momentum tensor, $T_{ab}$, normally can be defined. If the
background spacetime has a Killing field $k^a$, then $J^a = {T^a}_b
k^b$ is a locally conserved current. If $\Sigma$ is a Cauchy surface,
then $q = \int_\Sigma J^a d \Sigma_a$ defines a conserved quantity
associated with $k^a$; if $\Sigma$ is a timelike or null surface, then
$\int_\Sigma J^a d \Sigma_a$ has the interpretion of the flux of this
quantity through $\Sigma$.

However, in diffeomorphism covariant theories such as general
relativity, there is no notion of the local stress-energy tensor of
the gravitational field, so conserved quantities (which clearly must
include gravitational contributions) and their fluxes cannot be
defined by the above procedures, even when Killing fields are
present. Nevertheless, in general relativity, for asymptotically flat
spacetimes, conserved quantities associated with asymptotic symmetries
have been defined at spatial and null infinity.

A definition of mass-energy and radiated energy at null infinity,
${\cal I}$, was first given about 40 years ago by Trautman \cite{t}
and Bondi et al \cite{b}. This definition was arrived at via a
detailed study of the asymptotic behavior of the metric, and the main
justification advanced for this definition has been its agreement with
other notions of mass in some simple cases as well as the fact that
the radiated energy is always positive (see, e.g., \cite{g},
\cite{cjm} for further discussion of the justification for this
definition). A number of inequivalent definitions of quantities
associated with general (BMS) asymptotic symmetries at null infinity
have been proposed over the years, but it was not until the mid-1980's
that Dray and Streubel \cite{ds} gave a general definition that
appears to have fully satisfactory properties \cite{s}. This
definition generalized a definition of angular momentum given by
Penrose \cite{p} that was motivated by twistor theory.

In much of the body of work on defining ``conserved quantities'' at
null infinity, little contact has been made with the Hamiltonian
formulation of general relativity. An important exception is the work
of Ashtekar and Streubel \cite{as} (see also \cite{abr}), who noted
that BMS transformations correspond to canonical transformations on
the radiative phase space at ${\cal I}$. They identified the
Hamiltonian generating these canonical transformations as representing
the net flux of the ``conserved quantity'' through ${\cal I}$. They
then also obtained a local flux formula under some additional
assumptions not related to the canonical framework (in particular, by
their choice of topology they, in effect, imposed the condition that
the local flux formula contain no ``second derivative
terms''). However, they did not attempt to derive a local expression
for the ``conserved quantity'' itself within the Hamiltonian
framework, and, indeed, until the work of \cite{ds} and \cite{s}, it
was far from clear that, for arbitrary BMS generators, their flux
formula corresponds to a quantity that could be locally defined on
cross-sections of ${\cal I}$.

The status of the definition of ``conserved quantities'' at null
infinity contrasts sharply with the situation at spatial infinity,
where formulas for conserved quantities have been derived in a clear
and straightforward manner from the Hamiltonian formulation of general
relativity \cite{adm}, \cite{rt}. As will be reviewed in sections 2
and 3 below, for a diffeomorphism covariant theory derived from a
Lagrangian, if one is given a spacelike slice $\Sigma$ and a vector
field $\xi^a$ representing ``time evolution'', then the Hamiltonian
generating this time evolution---if it exists---must be purely a
``surface term'' when evaluated on solutions (i.e., ``on shell''). It
can be shown that if $\Sigma$ extends to spatial infinity in a
suitable manner and if $\xi^a$ is a suitable infinitesimal asymptotic
symmetry, then a Hamiltonian does exist (see ``case I'' of section 4
below). The value of this Hamiltonian ``on shell'' then can be
interpreted as being the conserved quantity conjugate to $\xi^a$. One
thereby directly obtains formulas for the ADM mass, momentum, and
angular momentum as limits as one approaches spatial infinity of
surface integrals over two-spheres.

It might seem natural to try a similar approach at null infinity: Let
$\Sigma$ be a spacelike slice which is asymptotically null in the
sense that in the unphysical spacetime its boundary is a
cross-section, ${\cal C}$, of ${\cal I}$. Let the vector field $\xi^a$
be an infinitesimal BMS asymptotic symmetry. Then, when evaluated on
solutions, the Hamiltonian generating this time evolution---if it
exists---must again be purely a ``surface term'' on $\Sigma$, i.e., it
must be expressible as an integral of a local expression over the
cross-section ${\cal C}$. This expression would then provide a natural
candidate for the value of the ``conserved quantity'' conjugate to
$\xi^a$ at ``time'' ${\cal C}$.

As we shall see in section 3 below, the above proposal works if
$\xi^a$ is everywhere tangent to ${\cal C}$. However, if $\xi^a$ fails
to be everywhere tangent to ${\cal C}$, then it is easy to show that
no Hamiltonian generating the time evolution exists. The
obstruction to defining a Hamiltonian arises directly from the
possibility that symplectic current can escape through ${\cal C}$.

The main purpose of this paper is to propose a general prescription
for defining ``conserved quantities'' in situations where a
Hamiltonian does not exist. This proposal consists of modifying the
equation that a Hamiltonian must satisfy via the addition of a
``correction term'' involving a symplectic potential that is required
to vanish whenever the background spacetime is stationary. If such a
symplectic potential exists and is unique---and if a suitable
``reference solution'' can be chosen to fix the arbitrary constant in
the definition of the ``conserved quantity''---we obtain a unique
prescription for defining a ``conserved quantity'' associated with any
infinitesimal asymptotic symmetry. In the case of asymptotically flat
spacetimes at null infinity in vacuum general relativity, we show in
section 5 that existence and uniqueness does hold, and that this
prescription yields the quantities previously obtained in \cite{ds}.

In section 2, we review some preliminary material on the
diffeomorphism covariant theories derived from a Lagrangian. In
section 3, we investigate the conditions under which a Hamiltonian
exists. In section 4, we present, in a very general setting, our
general proposal for the definition of ``conserved quantities''
associated with infinitesimal asymptotic symmetries. This general
proposal is then considered in the case of asymptotically flat
spacetimes at null infinity in general relativity in section 5, where
it is shown to yield the results of \cite{ds}. Some further
applications are briefly discussed in section 6.

\section{Preliminaries}

In this paper, we will follow closely both the conceptual framework
and the notational conventions of \cite {lw} and \cite{iw}. Further
details of most of what is discussed in this section can be found in those
references.

On an $n$-dimensional manifold, $M$, we consider a theory of dynamical
fields, collectively denoted $\phi$, which consist of a Lorentzian
metric, $g_{ab}$, together with other tensor fields, collectively
denoted as $\psi$. To proceed, we must define a space, ${\cal F}$, of
``kinematically allowed'' field configurations, $\phi = (g_{ab},
\psi)$ on $M$. A precise definition of ${\cal F}$ would involve the
specification of smoothness properties of $\phi$, as well as possible
additional restrictions on $g_{ab}$ (such as global hyperbolicity or
the requirement that a given foliation of $M$ by hypersurfaces be
spacelike) and asymptotic conditions on $\phi$ (such as the usual
asymptotic flatness conditions on fields at spatial and/or null
infinity in general relativity). The precise choice of ${\cal F}$ that
would be most suitable for one's purposes would depend upon the
specific theory and issues being considered. In this section and the
next section, we will merely assume that a suitable ${\cal F}$ has
been defined in such a way that the integrals occurring in the various
formulas below converge.  In section 4, we will impose a general set
of conditions on $\cal F$ that will ensure convergence of all relevant
integrals. In section 5, we will verify that asymptotically flat
spacetimes at null infinity in vacuum general relativity satisfy these
conditions.

We assume that the equations of motion of the theory arise from a
diffeomorphism covariant $n$-form Lagrangian density \cite{iw}
\begin{equation}
{\bf L} = {\bf L} \left( g_{ab}; R_{abcd}, \nabla_a R_{bcde},
...;\psi, \nabla_a \psi, ...\right)
\label{lag}
\end{equation}
where $\nabla_a$ denotes the derivative operator associated with
$g_{ab}$, $R_{abcd}$ denotes the Riemann curvature tensor of
$g_{ab}$. (An arbitrary (but finite) number of derivatives of
$R_{abcd}$ and $\psi$ are permitted to appear in ${\bf L}$.) Here and
below we use boldface letters to denote differential forms on
spacetime and, when we do so, we will suppress the spacetime indices
of these forms. Variation of ${\bf L}$ yields
\begin{equation}
\delta {\bf L} = {\bf E}(\phi) \delta \phi + d {\mbox{\boldmath
$\theta$}}(\phi, \delta \phi) .
\label{dL}
\end{equation}
where no derivatives of $\delta \phi$ appear in the first term on the
right side. The Euler-Lagrange equations of motion of the theory are
then simply ${\bf E} = 0$. Note that---when the variation is performed
under an integral sign---the term ${\mbox{\boldmath $\theta$}}$
corresponds to the boundary term that arises from the integrations by
parts needed to remove derivatives from $\delta \phi$. We require that
${\mbox{\boldmath $\theta$}}$ be locally constructed out of $\phi$ and
$\delta \phi$ in a covariant manner. This restricts the freedom in
the choice of ${\mbox{\boldmath $\theta$}}$ to\footnote{If we change
the Lagrangian by ${\bf L} \rightarrow {\bf L} + d {\bf K}$, the
equations of motion are unaffected. Under such change in the
Lagrangian, we have ${\mbox{\boldmath $\theta$}} \rightarrow
{\mbox{\boldmath $\theta$}} + \delta {\bf K}$. Thus, if such changes
in the Lagrangian are admitted, we will have this additional ambiguity
in ${\mbox{\boldmath $\theta$}}$. However, this ambiguity does not
affect the definition of the presymplectic current form (see
eq.(\ref{omega}) below) and will not affect our analysis.}
\begin{equation}
{\mbox{\boldmath $\theta$}} \rightarrow {\mbox{\boldmath $\theta$}} +
d {\bf Y}
\label{Y1}
\end{equation}
where ${\bf Y}$ is locally constructed out of $\phi$ and $\delta
\phi$ in a covariant manner. 

The presympletic current $(n-1)$-form, ${\mbox{\boldmath
$\omega$}}$---which is a local function of a field configuration,
$\phi$, and two linearized perturbations, $\delta_1 \phi$ and
$\delta_2 \phi$ off of $\phi$---is obtained by taking an
antisymmetrized variation of ${\mbox{\boldmath $\theta$}}$
\begin{equation}
{\mbox{\boldmath $\omega$}} (\phi, \delta_1 \phi, \delta_2 \phi) =
\delta_1 {\mbox{\boldmath $\theta$}} (\phi,\delta_2
\phi)-\delta_2{\mbox{\boldmath $\theta$}} (\phi,\delta_1 \phi)
\label{omega}
\end{equation}
On account of the ambiguity (\ref{Y1}) in the choice of 
${\mbox{\boldmath $\theta$}}$, we have the ambiguity
\begin{equation}
{\mbox{\boldmath $\omega$}} \rightarrow {\mbox{\boldmath $\omega$}} +
d [\delta_1 {\bf Y}(\phi, \delta_2 \phi) - \delta_2 {\bf Y}(\phi,
\delta_1 \phi)]
\label{Y2}
\end{equation}
in the choice of ${\mbox{\boldmath $\omega$}}$. 

Now let $\Sigma$ be a closed, embedded $(n-1)$-dimensional submanifold
without boundary; we will refer to $\Sigma$ as a {\it slice}.  The
presymplectic form, $\Omega_\Sigma$, associated with $\Sigma$ is a map
taking field configurations, $\phi$, together with a pairs of
linearized perturbations off of $\phi$, into the real numbers---i.e.,
it is a two-form on ${\cal F}$---defined by integrating\footnote{The
orientation of $\Sigma$ relative to the spacetime orientation
$\epsilon_{a_1...a_n}$ is chosen to be $v^{a_1} \epsilon_{a_1...a_n}$
where $v^a$ is a future-directed timelike vector.\label{orient1}}
${\mbox{\boldmath $\omega$}}$ over $\Sigma$,
\begin{equation}
\Omega_\Sigma (\phi, \delta_1 \phi, \delta_2 \phi) = \int_\Sigma
{\mbox{\boldmath $\omega$}}
\label{Omega}
\end{equation}
Although this definition depends, in general, upon the choice of
$\Sigma$, if $\delta_1 \phi$ and $\delta_2 \phi$ satisfy the
linearized field equations and $\Sigma$ is required to be a Cauchy
surface, then $\Omega_\Sigma$ does not depend upon the choice of $\Sigma$,
provided that $\Sigma$ is compact or suitable asymptotic conditions
are imposed on the dynamical fields \cite{lw}.  The ambiguity
(\ref{Y2}) in the choice of ${\mbox{\boldmath $\omega$}}$ gives rise
to the ambiguity
\begin{equation}
\Omega_\Sigma (\phi, \delta_1 \phi, \delta_2 \phi) \rightarrow \Omega_\Sigma (\phi,
\delta_1 \phi, \delta_2 \phi) + \int_{\partial \Sigma} [\delta_1 {\bf
Y}(\phi, \delta_2 \phi) - \delta_2 {\bf Y}(\phi, \delta_1 \phi)]
\label{Y3}
\end{equation}
in the presymplectic form $\Omega_\Sigma$. In this equation, by the
integral over $\partial \Sigma$, we mean a limiting process in which
the integral is first taken over the boundary, $\partial K$, of a
compact region, $K$, of $\Sigma$ (so that Stokes' theorem can be
applied\footnote{We choose the orientation of $\partial K$ to be the
one specified by Stokes' theorem, i.e., we dot the first index of the
orientation form on $K$ into an outward pointing
vector.\label{orient2}}), and then $K$ approaches all of $\Sigma$ in a
suitably specified manner. (Note that since $\Sigma$ is a slice, by
definition it does not have an actual boundary in the spacetime.)
Thus, for example, if $\Sigma$ is an asymptotically flat spacelike
slice in an asymptotically flat spacetime, the integral on the right
side of eq.(\ref{Y3}) would correspond to the integral over a
two-sphere on $\Sigma$ in the asymptotically flat region in the limit
as the radius of the two-sphere approaches infinity. Of course, the
right side of eq.(\ref{Y3}) will be well defined only if this limit
exists and is independent of any of the unspecified details of how the
compact region, $K$, approaches $\Sigma$. In section 4 below, we will
make some additional assumptions that will ensure that integrals over
``$\partial \Sigma$'' of certain quantities that we will consider are
well defined.

Given the presymplectic form, $\Omega_\Sigma$, we can factor ${\cal
F}$ by the orbits of the degeneracy subspaces of $\Omega_\Sigma$ to
construct a phase space, $\Gamma$, in the manner described in
\cite{lw}. This phase space acquires directly from the presymplectic
form $\Omega_\Sigma$ on ${\cal F}$ a nondegenerate symplectic form,
$\Omega$. One also obtains by this construction a natural projection
from ${\cal F}$ to $\Gamma$. Now, a complete vector field $\xi^a$ on
$M$ naturally induces the field variation ${\cal L}_\xi \phi$ on
fields $\phi \in {\cal F}$. If $\xi^a$ is such that ${\cal L}_\xi
\phi$ corresponds to a tangent field on ${\cal F}$ (i.e., if the
diffeomorphisms generated by $\xi^a$ map $\cal F$ into itself), then
we may view $\delta_\xi \phi = {\cal L}_\xi \phi$ as the dynamical
evolution vector field corresponding to the notion of ``time
translations'' defined by $\xi^a$. If, when restricted to the solution
submanifold\footnote{The solution submanifold, $\bar{{\cal F}}$, is
sometimes referred to as the ``covariant phase space'' \cite{abr}.},
$\bar{{\cal F}}$, of ${\cal F}$, this time evolution vector field on
${\cal F}$ consistently projects to phase space, then one has a notion
of time evolution associated with $\xi^a$ on the ``constraint
submanifold'', $\bar{\Gamma}$, of $\Gamma$, where $\bar{\Gamma}$ is
defined to be the image of $\bar{{\cal F}}$ under the projection of
${\cal F}$ to $\Gamma$. If this time evolution vector field on
$\bar{\Gamma}$ preserves the pullback to $\bar{\Gamma}$ of $\Omega$,
it will be generated by a Hamiltonian, $H_\xi$ \cite{lw}. (As argued
in the Appendix of \cite{lw}, this will be the case when $\Sigma$ is
compact; see section 3 below for some general results in the
noncompact case.)  Thus, this construction provides us with the notion
of a Hamiltonian, $H_\xi$, conjugate to a vector field $\xi^a$ on $M$.

However, a number of complications arise in the above
construction. In particular, in order to obtain a consistent
projection of ${\cal L}_\xi \phi$ from $\bar{{\cal F}}$ to
$\bar{\Gamma}$, it is necessary to choose $\xi^a$ to be ``field
dependent'', i.e., to depend upon $\phi$. As explained in \cite{lw},
this fact accounts for why, in a diffeomorphism covariant theory, the
Poisson bracket algebra of constraints does not naturally correspond
to the Lie algebra of infinitesimal diffeomorphisms. However, these
complications are not relevant to our present concerns. To avoid
dealing with them, we prefer to work on the original field
configuration space ${\cal F}$ with its (degenerate) presymplectic
form $\Omega_\Sigma$ rather than on the phase space $\Gamma$. The notion of a
Hamiltonian, $H_\xi$, on ${\cal F}$ can be defined as follows:

\medskip

\noindent
{\it Definition}: Consider a diffeomorphism covariant theory within
the above framework, with field configuration space ${\cal F}$ and
solution submanifold $\bar{\cal F}$.  Let $\xi^a$ be a vector field on
the spacetime manifold, $M$, let $\Sigma$ be a slice of $M$, and let
$\Omega_\Sigma$ denote the presymplectic form (\ref{Omega}). (If the
ambiguity (\ref{Y2}) in the choice of ${\mbox{\boldmath $\omega$}}$
gives rise to an ambiguity in $\Omega_\Sigma$ (see eq.(\ref{Y3})),
then we assume that a particular choice of $\Omega_\Sigma$ has been
made.)  Suppose that ${\cal F}$, $\xi^a$, and $\Sigma$ have been
chosen so that the integral $\int_\Sigma {\mbox{\boldmath
$\omega$}}(\phi, \delta \phi, {\cal L}_\xi \phi)$
converges for all $\phi \in \bar{\cal F}$ and all tangent vectors
$\delta \phi$ to $\bar{\cal F}$ at $\phi$. Then a function $H_\xi :
{\cal F} \rightarrow {\rm I} \! {\rm R}$ is said to be a {\em
Hamiltonian conjugate to $\xi^a$} on slice $\Sigma$ if for all $\phi
\in \bar{\cal F}$ and all field variations $\delta \phi$ tangent to
$\cal F$ (but not necessarily tangent to $\bar{\cal F}$) we have
\begin{equation}
\delta H_\xi = \Omega_\Sigma(\phi, \delta \phi, {\cal L}_\xi \phi) =
\int_\Sigma
{\mbox{\boldmath $\omega$}}(\phi, \delta \phi, {\cal L}_\xi \phi)
\label{H}
\end{equation}

\medskip

Note that if a Hamiltonian conjugate to $\xi^a$ on slice $\Sigma$
exists, then---assuming that $\bar{\cal F}$ is connected---its value
on $\bar{\cal F}$ is uniquely determined by eq.(\ref{H}) up to the
addition of an arbitrary constant. In many situations, this constant
can be fixed in a natural way by requiring $H_\xi$ to vanish for a
natural reference solution, such as Minkowski spacetime. On the other
hand, the value of $H_\xi$ off of $\bar{\cal F}$ is essentially
arbitrary, since eq.(\ref{H}) fixes only the ``field space gradient''
of $H_\xi$ in directions off of $\bar{\cal F}$ at points of $\bar{\cal
F}$.

If a Hamiltonian conjugate to $\xi^a$ on slice $\Sigma$ exists, then
its value provides a natural definition of a conserved quantity
associated with $\xi^a$ at ``time'' $\Sigma$. However, in many cases
of interest---such as occurs in general relativity when, say, $\xi^a$
is an asymptotic time translation and the slice $\Sigma$ goes to null
infinity---no Hamiltonian exists. In the next section, we shall
analyze the conditions under which a Hamiltonian exists. In
section 4, we shall propose a definition of the ``conserved
quantity'' conjugate to $\xi^a$ on a slice $\Sigma$ when no Hamiltonian
exists.

\section{Existence of a Hamiltonian}

When does a Hamiltonian conjugate to $\xi^a$ on slice $\Sigma$ exist?
To analyze this issue, it is very useful to introduce the Noether
current $(n-1)$-form associated with $\xi^a$, defined by
\begin{equation}
{\bf j} = {\mbox{\boldmath $\theta$}} (\phi, {\cal L}_\xi \phi) -
\xi \cdot {\bf L}
\label{j}
\end{equation}
where the ``$\cdot$'' denotes the contraction of the vector field
$\xi^a$ into the first index of the differential form ${\bf L}$. One
can show (see the appendix of \cite{iw2}) that for a diffeomorphism
covariant theory, ${\bf j}$ always can be written in the form
\begin{equation}
{\bf j} = d {\bf Q} + \xi^a {\bf {C}}_a ,
\label{Q}
\end{equation}
where ${\bf {C}}_a = 0$ when the equations of motion hold, i.e., ${\bf
{C}}_a$ corresponds to ``constraints'' of the theory. Equation
(\ref{Q}) defines the Noether charge $(n-2)$-form, $\bf Q$. It was
shown in \cite{iw} that the Noether charge always takes the form
\begin{equation}
{\bf Q} = {\bf X}^{ab}(\phi) \nabla_{[a} \xi_{b]} + {\bf U}_a (\phi)
\xi^a + {\bf V} (\phi, {\cal L}_\xi \phi) + d {\bf Z}(\phi, \xi) .
\label{Qform}
\end{equation}

{}From eqs.(\ref{dL}), (\ref{omega}), and (\ref{j}), it follows
immediately that for $\phi \in \bar{\cal F}$ but $\delta \phi$
arbitrary (i.e., $\delta \phi$ tangent to $\cal F$ but not necessarily
tangent to $\bar{\cal F}$), the variation of ${\bf j}$ satisfies
\begin{equation}
\delta {\bf j} = {\mbox{\boldmath $\omega$}} (\phi, \delta \phi, {\cal
L}_\xi \phi) + d(\xi \cdot {\mbox{\boldmath $\theta$}}) .
\label{dj}
\end{equation}
Thus, we obtain
\begin{equation}
{\mbox{\boldmath $\omega$}} (\phi, \delta \phi, {\cal L}_\xi \phi) =
\xi^a \delta {\bf {C}}_a + d (\delta {\bf Q}) - d(\xi \cdot
{\mbox{\boldmath $\theta$}}) .
\label{dj2}
\end{equation}
Consequently, if there exists a Hamiltonian, $H_\xi$, conjugate to
$\xi^a$ on $\Sigma$, then for all $\phi \in \bar{\cal F}$ and all $\delta
\phi$ it must satisfy the equation
\begin{equation}
\delta H_\xi = \int_\Sigma
\xi^a \delta {\bf {C}}_a + \int_{\partial \Sigma} [\delta {\bf Q} - \xi \cdot
{\mbox{\boldmath $\theta$}}]
\label{dh}
\end{equation}
where the integral over $\partial \Sigma$ has the meaning explained
below eq.(\ref{Y3}). Note that for field variations which are ``on
shell'', i.e., such that $\delta \phi$ satisfies the linearized
equations of motion, we have
\begin{equation}
\delta H_\xi = \int_{\partial \Sigma} [\delta {\bf Q} - \xi \cdot
{\mbox{\boldmath $\theta$}}] .
\label{hsurf}
\end{equation}
Consequently, if $H_\xi$ exists, it is given purely as a ``surface
term'' (i.e., an integral over $\partial \Sigma$) when evaluated on
$\bar{\cal F}$.

Equation (\ref{dh}) gives rise to an obvious necessary condition for
the existence of $H_\xi$: Let $\phi \in \bar{\cal F}$ (i.e., $\phi$ is
a solution to the field equations) and let $\delta_1 \phi$ and
$\delta_2 \phi$ be tangent to $\bar{\cal F}$ (i.e., $\delta_1 \phi$
and $\delta_2 \phi$ satisfy the linearized field equations). Let
$\phi(\lambda_1, \lambda_2)$ be a two-parameter family with $\phi(0,0)
= \phi$, $\partial \phi/\partial \lambda_1 (0,0) = \delta_1 \phi$, and
$\partial \phi/\partial \lambda_2 (0,0) = \delta_2 \phi$. Then if
eq.(\ref{dh}) holds, by equality of mixed partial derivatives, we must
have
\begin{eqnarray}
0 & = & (\delta_1 \delta_2 - \delta_2 \delta_1) H_\xi \nonumber\\ 
& = & - \int_{\partial \Sigma} \xi \cdot [\delta_1 {\mbox{\boldmath
$\theta$}}(\phi, \delta_2 \phi) - \delta_2 {\mbox{\boldmath
$\theta$}}(\phi, \delta_1 \phi)] \nonumber\\
& = &  - \int_{\partial \Sigma} \xi \cdot {\mbox{\boldmath
$\omega$}}(\phi, \delta_1 \phi, \delta_2 \phi)
\label{d2h}
\end{eqnarray}
Conversely, if eq.(\ref{d2h}) holds, then---assuming that $\bar{\cal
F}$ is simply connected (and has suitable differentiable
properties)---it will be possible to define $H_\xi$ on $\bar{\cal F}$
so that eq.(\ref{dh}) holds whenever $\delta \phi$ is tangent to
$\bar{\cal F}$. ({\it Proof}: On each connected component of
$\bar{\cal F}$ choose a ``reference solution'' $\phi_0 \in \bar{\cal
F}$ and define $H_\xi = 0$ at $\phi_0$. Let $\phi \in \bar{\cal F}$
and let $\phi(\lambda)$ for $\lambda \in [0,1]$ be a smooth,
one-parameter family of solutions that connects $\phi_0$ to
$\phi$. Define
\begin{equation}
H_\xi [\phi] = \int_0^1 d \lambda \int_{\partial \Sigma} [\delta {\bf
Q}(\lambda) - \xi \cdot {\mbox{\boldmath $\theta$}}(\lambda)] .
\label{hdef}
\end{equation}
This definition will be independent of the choice of path
$\phi(\lambda)$ when eq.(\ref{d2h}) holds since, by
simple-connectedness, any other path $\phi'(\lambda)$ will be
homotopic to $\phi(\lambda)$ and one can apply Stokes' theorem to the
two-dimensional submanifold spanned by this homotopy.)  However, if
$H_\xi$ is defined on $\bar{\cal F}$, there is no obstruction to
extending $H_\xi$ to ${\cal F}$ so that eq.(\ref{dh}) holds on
$\bar{\cal F}$ for all $\delta \phi$ tangent to $\cal F$ (i.e.,
including $\delta \phi$ that are not tangent to $\bar{\cal F}$), since
the additional content of that equation merely fixes the first
derivative of $H_\xi$ in the ``off shell'' directions of field
space. 

Therefore, the necessary and sufficient condition for the
existence of a Hamiltonian conjugate to $\xi^a$ on $\Sigma$ is that
for all solutions $\phi \in \bar{\cal F}$ and all pairs of linearized
solutions $\delta_1 \phi, \delta_2 \phi$ tangent to $\bar{\cal F}$, we
have
\begin{equation}
\int_{\partial \Sigma} \xi \cdot {\mbox{\boldmath
$\omega$}}(\phi, \delta_1 \phi, \delta_2 \phi) = 0 .
\label{Hexist}
\end{equation}
Note that since this condition refers only to the ``covariant phase
space'' $\bar{\cal F}$, we shall in the following restrict attention
to entirely $\bar{\cal F}$ and use eq.(\ref{hsurf}) for $H_\xi$ (even
though the ``off shell'' volume integral in eq.(\ref{dh}) is crucial
to justifying the interpretation of $H_\xi$ as the generator of
dynamics conjugate to $\xi^a$).

Note that there are two situations where eq.(\ref{Hexist}) will
automatically hold: (i) if the asymptotic conditions on $\phi$ are
such that ${\mbox{\boldmath $\omega$}}(\phi, \delta_1 \phi, \delta_2
\phi)$ goes to zero sufficiently rapidly that the integral of $\xi
\cdot {\mbox{\boldmath $\omega$}}$ over $\partial K$ vanishes in the
limit as $K$ approaches $\Sigma$; (ii) if $\xi^a$ is such that $K$ can
always be chosen so that $\xi^a$ is tangent to $\partial K$, since
then the pullback of $\xi \cdot {\mbox{\boldmath $\omega$}}$ to
$\partial K$ vanishes. In two these cases, a Hamiltonian conjugate to
$\xi^a$ will exist on $\Sigma$. However, if these conditions do not
hold, then in general no Hamiltonian will exist.

We turn, now, to giving a general prescription for defining ``conserved
quantities'', even when no Hamiltonian exists.

\section{General Definition of ``Conserved Quantities''}

In this section, we will propose a definition of conserved quantities
under very general assumptions about asymptotic conditions ``at
infinity''. We begin by specifying these assumptions.

We shall assume that the desired asymptotic conditions in the given
diffeomorphism covariant theory under consideration are specified by
attaching a boundary, ${\cal B}$, to the spacetime manifold, $M$, and
requiring certain limiting behavior of the dynamical fields, $\phi$,
as one approaches ${\cal B}$. We shall assume that ${\cal B}$ is an
$(n-1)$-dimensional manifold, so that $M \cup {\cal B}$ is an
$n$-dimensional manifold with boundary\footnote{The assumption that
${\cal B}$ is an $(n-1)$-dimensional manifold structure is not
essential in cases where ${\mbox{\boldmath $\omega$}}$ vanishes at
${\cal B}$ (see ``Case I'' below). In particular, there should be no
difficulty in extending our framework to definitions of asymptotic
flatness at spatial infinity in which ${\cal B}$ is comprised by a
single point \cite{ah}.}. In cases of interest, $M \cup {\cal B}$ will
be equipped with additional nondynamical structure (such as a
conformal factor on $M \cup {\cal B}$ or certain tensor fields on
$\cal B$) that will enter into the specification of the limiting
behavior of $\phi$ and thereby be part of the specification of the
field configuration space, ${\cal F}$, and the covariant phase space,
$\bar{\cal F}$. We will refer to such fixed, non-dynamical structure
as the ``universal background structure'' of $M \cup {\cal B}$.

We now state our two main assumptions concerning the asymptotic
conditions on the dynamical fields, $\phi$, and the asymptotic
behavior of the allowed hypersurfaces, $\Sigma$: (1) We assume that
${\cal F}$ has been defined so that for all $\phi \in \bar{\cal F}$
and for all $\delta_1 \phi, \delta_2 \phi$ tangent to $\bar{\cal F}$,
the $(n-1)$-form ${\mbox{\boldmath $\omega$}} (\phi, \delta_1 \phi,
\delta_2 \phi)$ defined on $M$ extends continuously\footnote{It should
be emphasized that we require that the full ${\mbox{\boldmath
$\omega$}}$ extend continuously to $\cal B$---not merely its pullback 
to hypersurfaces that approach $\cal B$.} to ${\cal B}$. (2)
We restrict consideration to slices, $\Sigma$, in the
``physical spacetime'', $M$, that extend smoothly to ${\cal B}$ in
the ``unphysical spacetime'', $M \cup {\cal B}$, such that this
extended hypersurface intersects ${\cal B}$ in a smooth
$(n-2)$-dimensional submanifold, denoted $\partial \Sigma$. Following
terminology commonly used for null infinity, we shall refer to
$\partial \Sigma$ as a ``cross-section'' of ${\cal B}$.
We also shall assume that $\Sigma \cup \partial \Sigma$ is
compact---although it would be straightforward to weaken this
assumption considerably, since only the behavior of $\Sigma$ near
${\cal B}$ is relevant to our considerations.

An important immediate consequence of the above two assumptions is
that the integral (\ref{Omega}) defining $\Omega_\Sigma$ always converges,
since it can be expressed as the integral of a continuous $(n-1)$-form
over the compact $(n-1)$-dimensional hypersurface $\Sigma \cup
\partial \Sigma$.

We turn, now, to the definition of infinitesimal asymptotic
symmetries. Let $\xi^a$ be a complete vector field on $M \cup {\cal
B}$ (so that, in particular, $\xi^a$ is tangent to $\cal B$ on $\cal
B$). We say that $\xi^a$ is a {\it representative of an infinitesimal
asymptotic symmetry} if its associated one-parameter group of
diffeomorphisms maps $\bar{\cal F}$ into $\bar{\cal F}$, i.e., if it
preserves the asymptotic conditions specified in the definition of
$\bar{\cal F}$. Equivalently, $\xi^a$ is a representative of an
infinitesimal asymptotic symmetry if ${\cal L}_\xi \phi$ (which
automatically satisfies the linearized field equations \cite{lw})
satisfies all of the asymptotic conditions on linearized solutions arising
from the asymptotic conditions imposed upon $\phi \in \bar{\cal F}$,
i.e., if ${\cal L}_\xi \phi$ corresponds to a vector tangent to
$\bar{\cal F}$.

If $\xi^a$ is a representative of an infinitesimal asymptotic
symmetry, then the integral appearing on the right side of
eq.(\ref{hsurf}), namely
\begin{equation}
I = \int_{\partial \Sigma} [\delta {\bf Q} - \xi \cdot
{\mbox{\boldmath $\theta$}}] 
\label{I}
\end{equation}
always is well defined via the limiting proceedure described below
eq.(\ref{Y3}), and, indeed, $I$ depends only on the cross-section
$\partial \Sigma$ of ${\cal B}$, not on $\Sigma$. To see
this\footnote{A similar argument has previously been given to show
that the ``linkage formulas'' are well defined (see, \cite{tw},
\cite{gw}).}, let $K_i$ be a nested sequence of compact
subsets of $\Sigma$ such that $\partial K_i$ approaches $\partial
\Sigma$, and let
\begin{equation}
I_i = \int_{\partial K_i} [\delta {\bf Q} - \xi \cdot
{\mbox{\boldmath $\theta$}}] ,
\label{Ii}
\end{equation}
Then, since ``on shell'' we have
\begin{equation}
{\mbox{\boldmath $\omega$}} (\phi, \delta \phi, {\cal L}_\xi \phi) =
d [\delta {\bf Q} - \xi \cdot {\mbox{\boldmath $\theta$}}]
\label{omQ}
\end{equation}
(see eq.(\ref{dj2}) above) we have by Stokes' theorem for $i \geq j$,
\begin{equation}
I_i - I_j = \int_{\Sigma_{ij}}d [\delta {\bf Q} - \xi \cdot
{\mbox{\boldmath $\theta$}}] = \int_{\Sigma_{ij}}
{\mbox{\boldmath $\omega$}}(\phi, \delta \phi, {\cal L}_\xi \phi)
\label{Iij}
\end{equation}
where $\Sigma_{ij}$ denotes $K_i \setminus K_j$, i.e., the portion of
$\Sigma$ lying between $\partial K_i$ and $\partial K_j$.  As a direct
consequence of our assumptions that ${\mbox{\boldmath $\omega$}}$
extends continuously to $\cal B$ and that $\Sigma \cup \partial
\Sigma$ is compact, it follows that $\{I_i\}$ is a Cauchy sequence,
and hence it has a well defined limit, $I$, as $i \rightarrow
\infty$. Note that this limit always exists despite the fact that
there is no guarantee that the differential forms ${\bf Q}$ or
${\mbox{\boldmath $\theta$}}$ themselves extend continuously to ${\cal
B}$. A similar argument establishes that this limit is independent of
$\Sigma$, i.e., for a slice $\tilde{\Sigma}$ such that $\partial
\tilde{\Sigma}$ = $\partial \Sigma$, a similarly defined sequence
$\{\tilde{I}_i \}$ of integrals on $\tilde{\Sigma}$ will also converge
to $I$.

Let $\xi^a$ and $\xi'^a$ be representatives of infinitesimal
asymptotic symmetries. We say that $\xi^a$ is {\it equivalent} to
$\xi'^a$ if they coincide on ${\cal B}$ and if, for all $\phi \in
\bar{\cal F}$, $\delta \phi$ tangent to $\bar{\cal F}$, and for all
$\partial \Sigma$ on ${\cal B}$, we have $I = I'$, where $I$ is given
by eq.(\ref{I}) and $I'$ is given by the same expression with $\xi^a$
replaced by $\xi'^a$. The {\it infinitesimal asymptotic symmetries} of
the theory are then comprised by the equivalence classes of the
representatives of the infinitesimal asymptotic symmetries.

Now consider an infinitesimal asymptotic symmetry, represented by the
vector field $\xi^a$, and let $\Sigma$ be a slice in the
spacetime with boundary $\partial \Sigma$ on ${\cal B}$. We would like
to define a conserved quantity $H_\xi: \bar{\cal F} \rightarrow {\rm
I} \! {\rm R}$ associated with $\xi^a$ at ``time'' $\Sigma$ via
eq.(\ref{hsurf}). As we have seen above, the right side of
eq.(\ref{hsurf}) is well defined under our asymptotic assumptions,
but, as discussed in the previous section, in general, there does not
exist an $H_\xi$ which satisfies this equation. The analysis naturally
breaks up into the following two cases:

\medskip
\noindent
{\bf Case I}: Suppose that the continuous extension of
${\mbox{\boldmath $\omega$}}$ to ${\cal B}$ has vanishing pullback to
${\cal B}$. Then by eq.(\ref{Hexist}), $H_\xi$ exists for all infinitesimal
asymptotic symmetries (assuming that $\bar{\cal F}$ is simply
connected and has suitable differentiable properties) and is
independent of the choice of representative $\xi^a$. Furthermore, if
$\partial \Sigma_1$ and $\partial \Sigma_2$ are cross-sections of
${\cal B}$ that bound a region ${\cal B}_{12} \subset {\cal B}$, we
have\footnote{We define the orientation of $\cal B$ to be that
obtained by dotting the first index of the orientation of $M$ into an
outward pointing vector. The orientation of $\partial \Sigma$ was
previously specified in footnotes \ref{orient1} and \ref{orient2}. The
signs in eq.(\ref{Hnocons}) to correspond to the case where $\partial
\Sigma_2$ lies to the future of $\partial \Sigma_1$.\label{orient4}}
by eqs.(\ref{hsurf}) and (\ref{omQ})
\begin{equation}
\delta H_\xi|_{\partial \Sigma_2} - \delta H_\xi|_{\partial \Sigma_1} =
- \int_{{\cal B}_{12}} {\mbox{\boldmath $\omega$}}(\phi, \delta \phi,
{\cal L}_\xi \phi) = 0
\label{Hcons}
\end{equation}
Thus, $\delta H_\xi$ is independent of choice of cross-section within
the same homology class. If the arbitrary constant (for each
cross-section) in $H_\xi$ is fixed in such a way that there is a
``reference solution'' for which $H_\xi = 0$ on all cross-sections
(see below), then on all solutions $H_\xi$ will be independent of
choice of cross-section within the same homology class. Thus, in this
case, not only does $H_\xi$ exist, but it truly corresponds to a
conserved quantity, i.e., its value is independent of ``time'',
$\Sigma$.

\medskip

\noindent
{\bf Case II}: Suppose that the continuous extension of
${\mbox{\boldmath $\omega$}}$ to ${\cal B}$ does not, in general, have
vanishing pullback to ${\cal B}$. Then, in general, there does not
exist an $H_\xi$ satisfying eq.(\ref{hsurf}). One exception is the
case where $\xi^a$ and $\partial \Sigma$ are such that $\xi^a$ is
everywhere tangent to $\partial \Sigma$. In this case, if $\xi^a$ is
tangent to cross-sections $\partial \Sigma_1$ and $\partial \Sigma_2$,
we have
\begin{equation}
\delta H_\xi|_{\partial \Sigma_2} - \delta H_\xi|_{\partial \Sigma_1}
= - \int_{{\cal B}_{12}} {\mbox{\boldmath $\omega$}}(\phi, \delta \phi,
{\cal L}_\xi \phi)
\label{Hnocons}
\end{equation}
Since the right side of this equation is nonvanishing in general, we
see that even when $\xi^a$ is tangent to cross-sections so that
$H_\xi$ exists, $H_\xi$ will not be conserved.

\medskip

Case I arises in general relativity for spacetimes which are
asymptotically flat at spatial infinity as defined in \cite{ar}, and our
prescription for defining $H_\xi$ corresponds to that given in
\cite{adm} and \cite{rt}; see \cite{iw} for the explicit details of
how eq.(\ref{hsurf}) gives rise to the usual expression for ADM mass
when $\xi^a$ is an asymptotic time translation. As we shall discuss in
detail in the next section, case II arises in general relativity for
spacetimes which are asymptotically flat at null infinity.

The main purpose of this paper is to provide a general definition of a
``conserved quantity'' conjugate to an arbitrary infinitesimal
asymptotic symmetry $\xi^a$ in case II. In the following, we will
restrict attention to this case, and we will denote the quantity we
seek as ${\cal H}_\xi$ to distinguish it from a true Hamiltonian
$H_\xi$. As we have seen, in this case an attempt to define ${\cal
H}_\xi$ by eq.(\ref{hsurf}) fails the consistency check (\ref{d2h})
and thus does not define any quantity. However, consider the following
simple modification of eq.(\ref{hsurf}): On ${\cal B}$, let
${\mbox{\boldmath $\Theta$}}$ be a symplectic potential for the
pullback, $\bar{\mbox{\boldmath $\omega$}}$, of the (extension of the)
symplectic current form ${\mbox{\boldmath $\omega$}}$ to ${\cal B}$,
so that on ${\cal B}$ we have for all $\phi \in \bar{\cal F}$ and
$\delta_1 \phi$, $\delta_2 \phi$ tangent to $\bar{\cal F}$
\begin{equation}
\bar{\mbox{\boldmath $\omega$}}(\phi, \delta_1 \phi, \delta_2 \phi) =
\delta_1 {\mbox{\boldmath $\Theta$}}(\phi, \delta_2 \phi) -
\delta_2 {\mbox{\boldmath $\Theta$}}(\phi, \delta_1 \phi)
\label{Theta}
\end{equation}
We require that ${\mbox{\boldmath $\Theta$}}$ be locally
constructed\footnote{More precisely, by ``locally constructed'' we
mean the following: Suppose that $\chi : M \cup {\cal B} \rightarrow M
\cup {\cal B}$ is a diffeomorphism which preserves the universal
background structure. Suppose $(\phi, \delta \phi)$ and $(\phi',
\delta \phi')$ are such that there exists an open (in $M \cup {\cal
B}$) neighborhood, $\cal O$, of $p \in {\cal B}$ such that for all $x
\in M \cap \cal O$ we have $\phi = \chi_* \phi'$ and $\delta \phi =
\chi_* \delta \phi'$, where $\chi_*$ denotes the pullback map on
tensor fields associated with the diffeomorphism $\chi$. Then we
require that at $p$ we have ${\mbox{\boldmath $\Theta$}} = \chi_*
{\mbox{\boldmath $\Theta$}}'$.\label{loc}} out of the dynamical
fields, $\phi$, and their derivatives (or limits of such quantities to
${\cal B}$) as well as any fields present in the ``universal
background structure''. In the case where $\bf L$ (and, hence
${\mbox{\boldmath $\omega$}}$) is an analytic function\footnote{The
condition that $\bf L$ be an analytic function of its variables (as
occurs in essentially all theories ever seriously considered) has
nothing to do with any smoothness or analyticity conditions concerning
the behavior of the dynamical fields themselves on $M$. We do not
impose any analyticity conditions on the dynamical fields.} of its
variables (see eq.(\ref{lag})), we also require that ${\mbox{\boldmath
$\Theta$}}$ depend analytically on the dynamical fields; more
precisely, if $\phi(\lambda)$ is a one-parameter family of fields on
$M$ that depends analytically on $\lambda$ and satisfies suitable
uniformity conditions\footnote{For the case of asymptotically flat
spacetimes at null infinity, $\cal I$, a suitable uniformity condition
would be to require the unphysical fields to vary analytically with
$\lambda$ at $\cal I$.} near $\cal B$, we require that the
corresponding ${\mbox{\boldmath $\Theta$}}(\lambda)$ also depends
analytically on $\lambda$. If any arbitrary choices are made in the
specification of the background structure (such as a choice of
conformal factor in the definition of null infinity in general
relativity), then we demand that ${\mbox{\boldmath $\Theta$}}$ be
independent of such choices (so, in particular, in the case of null
infinity, ${\mbox{\boldmath $\Theta$}}$ is required to be conformally
invariant). Our proposal is the following: Let ${\cal H}_\xi$
satisfy\footnote{Here
it should be noted that the new term on the right side of this
equation is an ordinary integral over the surface $\partial \Sigma$ of
${\cal B}$, whereas, as explained above, the first term in general is
defined only as an asymptotic limit.}
\begin{equation}
\delta {\cal H}_\xi = \int_{\partial \Sigma} [\delta {\bf Q} - \xi \cdot
{\mbox{\boldmath $\theta$}}] + \int_{\partial \Sigma} \xi \cdot
{\mbox{\boldmath $\Theta$}}
\label{hsurf2}
\end{equation}
Then it is easily seen that this formula satisfies the consistency
check (\ref{d2h}) and, thus, defines a ``conserved quantity'' ${\cal
H}_\xi$ up to an arbitrary constant.  Finally, let this arbitrary
constant be fixed by requiring that ${\cal H}_\xi$ vanish (for all
infinitesimal asymptotic symmetries $\xi^a$ and all cross-sections
$\partial \Sigma$) on a suitably chosen ``reference solution'' $\phi_0
\in \bar{\cal F}$. We will specify below the necessary conditions that
must be satisfied by $\phi_0$.

However, the above proposal fails to define a unique prescription
because the choice of symplectic potential ${\mbox{\boldmath
$\Theta$}}$ is ambiguous up to\footnote{Note that the ambiguity in
${\mbox{\boldmath $\Theta$}}$ is of an entirely different nature than
the ambiguity (\ref{Y1}) in ${\mbox{\boldmath $\theta$}}$. The
quantity ${\mbox{\boldmath $\theta$}}$ is defined from the Lagrangian
${\bf L}$ ({\it before} ${\mbox{\boldmath $\omega$}}$ has been
defined) and its ambiguity arises from eq.(\ref{dL}). The quantity
${\mbox{\boldmath $\Theta$}}$ is defined from ${\mbox{\boldmath
$\omega$}}$ and its ambiguity arises from eq.(\ref{Theta}).}
\begin{equation}
{\mbox{\boldmath $\Theta$}}(\phi, \delta \phi) \rightarrow
{\mbox{\boldmath $\Theta$}}(\phi, \delta \phi) + \delta {\bf W}(\phi)
\label{W}
\end{equation}
where ${\bf W}$ is an $(n-1)$-form on ${\cal B}$ locally constructed
out of the dynamical fields $\phi$ as well as the universal background
structure defined on ${\cal B}$, with ${\bf W}$ independent of any
arbitrary choices made in the specification of the background
structure.  Thus, in order to obtain a prescription which defines
${\cal H}_\xi$, we must specify an additional condition or conditions
which uniquely select ${\mbox{\boldmath $\Theta$}}$.

An additional requirement on ${\mbox{\boldmath $\Theta$}}$ can be
motivated as follows. We have already seen from eq.(\ref{Hnocons})
above that ${\cal H}_\xi$ cannot, in general, be conserved, i.e.,
there must be a nonzero flux, ${\bf F}_\xi$, on ${\cal B}$ associated
with this ``conserved quantity''. This is to be expected on account of
the possible presence of radiation at ${\cal B}$. However, it seems
natural to demand that ${\bf F}_\xi$ vanish (and, thus, that ${\cal H}_\xi$
be conserved) in the case where no radiation is present at ${\cal
B}$. Such a case should occur when $\phi$ is a stationary solution,
i.e., when there exists a nonzero infinitesimal asymptotic symmetry
represented by an exact symmetry $t^a$---so that ${\cal L}_t \phi =
0$ in $M$---and $t^a$ is timelike in $M$ in a neighborhood of ${\cal
B}$. Hence, we wish to require that ${\bf F}_\xi$ vanish on ${\cal B}$
for all $\xi^a$ for stationary solutions. 

To see what condition on ${\mbox{\boldmath $\Theta$}}$ will ensure
that this holds, we note that from eq.(\ref{hsurf2}), it
follows immediately that
\begin{equation}
\delta {\cal H}_\xi|_{\partial \Sigma_2} - \delta {\cal
H}_\xi|_{\partial \Sigma_1} = - \int_{{\cal B}_{12}} \delta {\bf
F}_\xi
\label{F1}
\end{equation}
where the variation of the flux $(n-1)$-form, ${\bf F}_\xi$, on ${\cal
B}$ is given by
\begin{equation}
\delta {\bf F}_\xi = \bar{\mbox{\boldmath $\omega$}}(\phi, \delta \phi,
{\cal L}_\xi \phi) + d [\xi \cdot {\mbox{\boldmath $\Theta$}}(\phi,
\delta \phi)]
\label{F2}
\end{equation}
Here the first term in this equation arises from taking `$d$'
of the integrand of the first term in eq.(\ref{hsurf2}) (using
eq.(\ref{omQ}) above), whereas the second term is just the `$d$'
of the integrand of the second term in eq.(\ref{hsurf2}). 
However, we have
\begin{eqnarray}
d [\xi \cdot {\mbox{\boldmath $\Theta$}}(\phi, \delta \phi)] & = & {\cal
L}_\xi {\mbox{\boldmath $\Theta$}}(\phi, \delta \phi) \nonumber \\ 
& = & - \bar{\mbox{\boldmath $\omega$}}(\phi, \delta \phi, {\cal L}_\xi \phi)
+ \delta {\mbox{\boldmath $\Theta$}}(\phi,{\cal L}_\xi \phi)
\label{F3}
\end{eqnarray}
Thus, we obtain
\begin{equation}
\delta {\bf F}_\xi = \delta {\mbox{\boldmath $\Theta$}}(\phi,{\cal L}_\xi \phi)
\label{F4}
\end{equation}

We now impose the requirement that ${\mbox{\boldmath $\Theta$}}(\phi,
\delta \phi)$ vanish whenever $\phi$ is stationary (even when $\delta
\phi$ is non-stationary). We also explicitly assume that the reference
solution, $\phi_0$, (on which ${\cal H}_\xi$ vanishes for all cross-sections
and hence ${\bf F}_\xi = 0$) is stationary. Since both
${\mbox{\boldmath $\Theta$}}$ and ${\bf F}_\xi$ vanish on $\phi_0$, we
obtain from eq.(\ref{F4}) the remarkably simple formula
\begin{equation}
{\bf F}_\xi = {\mbox{\boldmath $\Theta$}}(\phi,{\cal L}_\xi \phi)
\label{F5}
\end{equation}
It then follows immediately (as a consequence of our choice of
${\mbox{\boldmath $\Theta$}}$) that ${\bf F}_\xi$ vanishes (for all
$\xi$) on stationary solutions, as we desired. Equation (\ref{F5})
also implies an additional desirable property of ${\bf F}_\xi$: We
have ${\bf F}_\xi = 0$ whenever $\xi^a$ is an exact symmetry---i.e.,
whenever ${\cal L}_\xi \phi = 0$---regardless of whether radiation may
be present.

If a symplectic potential ${\mbox{\boldmath $\Theta$}}$ satisfying our
above condition exists and is unique, then eq.(\ref{hsurf2}) together
with the requirement that ${\cal H}_\xi$ vanish (for all
cross-sections and all $\xi^a$) on a particular, specified solution,
$\phi_0$, uniquely determines ${\cal H}_\xi$. However, there remains a
potential difficulty in specifying $\phi_0$: If $\phi_0 \in \bar{\cal F}$ then
we also have $\psi_* \phi_0 \in \bar{\cal F}$, where $\psi: M \cup
{\cal B} \rightarrow M \cup {\cal B}$ is any diffeomorphism generated
by a representative of an infinitesimal asymptotic symmetry. Since we
have no meaningful way of distinguishing between $\phi_0$ and $\psi_*
\phi_0$, if we demand that ${\cal H}_\xi$ vanish on $\phi_0$ we must
also demand that it vanish on $\psi_* \phi_0$. However, this
overdetermines ${\cal H}_\xi$ (so that no solution exists) unless the
following consistency condition holds: Let $\eta^a$ be a
representative of an infinitesimal asymptotic symmetry and consider
the field variation about $\phi_0$ given by $\delta \phi = {\cal
L}_\eta \phi_0$. Since this corresponds to the action of an
infinitesimal asymptotic symmetry on $\phi_0$, under this field
variation we must have $\delta {\cal H}_\xi = 0$. On the other hand,
$\delta {\cal H}_\xi$ is specified by eq.(\ref{hsurf2}). Since under
this field variation we have
\begin{equation}
\delta {\bf Q}[\xi] = {\cal L}_\eta {\bf Q}[\xi] - {\bf Q}[{\cal
L}_\eta \xi]
\label{deltaQ}
\end{equation}
and since, by assumption, ${\mbox{\boldmath $\Theta$}}$ vanishes at
$\phi_0$, we find that the consistency requirement on $\phi_0$ is that
for all representatives $\xi^a$ and $\eta^a$ of infinitesimal
asymptotic symmetries and for all cross-sections $\partial \Sigma$, we
must have
\begin{equation}
0 = \int_{\partial \Sigma} \{{\cal L}_\eta {\bf Q}[\xi] - {\bf
Q}[{\cal L}_\eta \xi] - \xi \cdot {\mbox{\boldmath
$\theta$}}(\phi_0,{\cal L}_\eta \phi_0) \}
\label{consis}
\end{equation}
From eqs.(\ref{omQ}) and (\ref{Theta}) together with the vanishing of
${\mbox{\boldmath $\Theta$}}$ at $\phi_0$, it follows that the right
side of eq.(\ref{consis}) is independent of cross-section and thus
need only be checked at one cross-section. In addition,
eq.(\ref{consis}) manifestly holds when $\eta^a$ is an exact symmetry
of $\phi_0$---i.e., when ${\cal L}_\eta \phi_0 = 0$---since $\delta
\phi = 0$ in that case. Using
\begin{equation}
{\cal L}_\eta {\bf Q} = d (\eta \cdot {\bf Q}) + \eta \cdot d {\bf Q}
= d (\eta \cdot {\bf Q}) + \eta \cdot {\bf j}
\label{lieid}
\end{equation}
together with eq.(\ref{j}), we may rewrite eq.(\ref{consis}) in the form
\begin{equation}
0 = \int_{\partial \Sigma} \{ \eta \cdot {\mbox{\boldmath
$\theta$}}(\phi_0,{\cal L}_\xi \phi_0) - \xi \cdot {\mbox{\boldmath
$\theta$}}(\phi_0,{\cal L}_\eta \phi_0) - \eta \cdot (\xi \cdot {\bf
L}) - {\bf Q}[{\cal L}_\eta \xi] \}
\label{consis2}
\end{equation}
(Here, the integral over $\partial \Sigma$ is to be interpreted as an
asymptotic limit, with the limit guaranteed to exist by the argument
given above. If ${\bf L}$ extends continuously to ${\cal B}$ then the
term $\eta \cdot (\xi \cdot {\bf L})$ makes no contribution to the
integral since both $\eta^a$ and $\xi^a$ are tangent to $\cal B$.)
Since eq.(\ref{consis2}) is manifestly antisymmetric in $\eta^a$ and
$\xi^a$, it follows that the consistency condition also is
automatically satisfied whenever $\xi^a$ is an exact symmetry of
$\phi_0$. However, if both $\eta^a$ and $\xi^a$ are asymptotic
symmetries that are not exact symmetries of $\phi_0$, then
eq.(\ref{consis}) (or, equivalently, eq.(\ref{consis2})) yields a
nontrivial condition that must be satisfied by $\phi_0$.

To summarize, we propose the following prescription for defining
``conserved quantities'' in case II: Let ${\mbox{\boldmath $\Theta$}}$
be a symplectic potential on ${\cal B}$ (see eq.(\ref{Theta}) above)
which is locally constructed out of the dynamical fields and
background structure (and is an analytic function of the dynamical
fields when $\bf L$ is analytic), is independent of any arbitrary
choices made in specifying the background structure, and is such that
${\mbox{\boldmath $\Theta$}}(\phi, \delta \phi)$ vanishes for all
$\delta \phi$ tangent to $\bar{\cal F}$ whenever $\phi \in \bar{\cal
F}$ is stationary. (If it exists, such a ${\mbox{\boldmath $\Theta$}}$
is unique up to addition of a term $\delta {\bf W}$ where ${\bf W}$ is
locally constructed out of the dynamical fields and background
structure (and is analytic in the dynamical fields when $\bf L$ is
analytic), is independent of any arbitrary choices made in specifying
the background structure, and is such that $\delta {\bf W}$ vanishes
for all $\delta \phi$ whenever $\phi$ is stationary.) Let $\phi_0$ be
a stationary solution that satisfies eq.(\ref{consis}) (or,
equivalently, eq.(\ref{consis2})) for all infinitesimal asymptotic
symmetries $\eta^a$ and $\xi^a$. Then we define ${\cal H}_\xi$ by
eq.(\ref{hsurf2}) together with the requirement that ${\cal H}_\xi$
vanish on $\phi_0$. To the extent that a ${\mbox{\boldmath $\Theta$}}$
satisfying the above requirements exists and is unique, and to the
extent that a stationary $\phi_0$ satisfying (\ref{consis}) exists,
this defines a prescription for defining ``conserved quantities''
associated with asymptotic symmetries. This prescription automatically
gives rise to the flux formula (\ref{F5}), so that the flux vanishes
whenever $\phi$ is stationary or $\xi^a$ is an exact symmetry.

In the next section, we analyze what this general prescription yields
for the case of asymptotically flat spacetimes at null infinity in
vacuum general relativity.

\section{``Conserved Quantities'' at Null Infinity in General Relativity}

In vacuum general relativity, the manifold $M$ is taken to be
$4$-dimensional and the only dynamical field, $\phi$, is the spacetime
metric, $g_{ab}$. We shall write the varied field as
\begin{equation}
\gamma_{ab} \equiv \delta g_{ab}
\label{gamma}
\end{equation}
The Einstein-Hilbert Lagrangian of general relativity is
\begin{equation}
{\bf L} = \frac{1}{16 \pi} R {\mbox{\boldmath $\epsilon$}}
\label{Lgr}
\end{equation}
where $R$ denotes the scalar curvature of $g_{ab}$ and
${\mbox{\boldmath $\epsilon$}}$ is the spacetime volume form
associated with $g_{ab}$.  The presymplectic potential $3$-form
${\mbox{\boldmath $\theta$}}$ is given by
\begin{equation}
\theta_{abc} = \frac{1}{16 \pi} \epsilon_{dabc} v^d
\label{thetagr}
\end{equation}
where
\begin{equation}
v^a = g^{ae} g^{fh} [\nabla_f \gamma_{eh} - \nabla_e \gamma_{fh}]
\label{v}
\end{equation}
where $\nabla_a$ is the derivative operator associated with $g_{ab}$.
The corresponding presymplectic current $3$-form is \cite{bw}
\begin{equation}
\omega_{abc} = \frac{1}{16 \pi} \epsilon_{dabc} w^d
\label{omegagr}
\end{equation}
where
\begin{equation}
w^a = P^{abcdef}[\gamma_{2bc} \nabla_d \gamma_{1ef} - \gamma_{1bc}
\nabla_d \gamma_{2ef} ]
\label{w}
\end{equation}
with
\begin{equation}
P^{abcdef} = g^{ae}g^{fb}g^{cd} - \frac{1}{2}g^{ad}g^{be}g^{fc}
- \frac{1}{2}g^{ab}g^{cd}g^{ef} - \frac{1}{2}g^{bc}g^{ae}g^{fd} +
\frac{1}{2}g^{bc}g^{ad}g^{ef}
\label{P}
\end{equation}
Finally, the Noether charge $2$-form associated with a vector field
$\xi^a$ is given by \cite{iw}
\begin{equation}
Q_{ab}[\xi] = - \frac{1}{16 \pi} \epsilon_{abcd} \nabla^c \xi^d
\label{Qgr}
\end{equation}

We wish to consider spacetimes that are asymptotically flat at future
and/or past null infinity. For definiteness, we will consider future
null infinity. (Sign changes would occur in several formulas when we
consider past null infinity on account of our orientation convention
on $\cal B$ (see footnote \ref{orient4}).) We denote future null
infinity by ${\cal I}$ and adopt the standard definition of asymptotic
flatness there (see, e.g., \cite{w}). The key ingredient of this
definition is that there exist a smooth\footnote{The requirement of
smoothness could be weakened considerably without affecting our
analysis.}  metric, $\tilde{g}_{ab}$, on $M \cup {\cal I}$ and a
smooth function, $\Omega$, on $M \cup {\cal I}$ such that $\Omega > 0$
on $M$, $\Omega = 0$ on ${\cal I}$, and $\tilde{\nabla}_a \Omega$ is
null\footnote{For solutions to the vacuum field equations, it follows
from the fact that $\Omega = 0$ on ${\cal I}$ that $\tilde{\nabla}_a
\Omega$ is null on ${\cal I}$ in the metric $\tilde{g}_{ab}$.} and
nonvanishing everywhere on ${\cal I}$, and such that throughout $M$ we have
\begin{equation}
\tilde{g}_{ab} = \Omega^2 g_{ab}
\label{gtilde}
\end{equation}
We also assume that $\cal I$ has topology ${\rm S}^2 \times {\rm I} \!
{\rm R}$.  In the following all indices will be raised and lowered
using the ``unphysical metric'', $\tilde{g}_{ab}$.  We write
\begin{equation}
n_a = \tilde{\nabla}_a \Omega
\label{n}
\end{equation}
(Here $\tilde{\nabla}_a$ denotes the derivative operator associated
with $\tilde{g}_{ab}$, although, of course, since $\Omega$ is a
scalar, $\nabla_a \Omega$ is independent of the choice of derivative
operator.) We may use the freedom $\Omega \rightarrow \omega \Omega$
with $\omega$ a smooth, strictly positive function on $M \cup {\cal
I}$ to assume, without loss of generality, that the Bondi condition
\begin{equation}
\tilde{\nabla}_a n_b |_{\cal I} = 0
\label{bondi}
\end{equation}
holds. An immediate consequence of eq.(\ref{bondi}) is that on ${\cal
I}$ we have $\tilde{\nabla}_a (n^b n_b) = 2 n^b \tilde{\nabla}_a n_b =
0$, so in the Bondi gauge
\begin{equation}
n^a n_a = O(\Omega^2)
\label{n2}
\end{equation}
Without loss of generality (see, e.g., \cite{gw}), we also may assume
that the conformal factor, $\Omega$, on $M \cup {\cal I}$ and the
unphysical metric, $\tilde{g}_{ab}$, on ${\cal I}$ are universal
quantities, i.e., they may be assumed to be independent of the
physical metric, $g_{ab}$, on $M$. Without loss of generality, we may
(by use of freedom remaining in the choice of $\Omega$) take the
universal unphysical metric $\tilde{g}^0_{ab}$, on ${\cal I}$ to be
such that the induced spatial metric on all cross-sections of ${\cal
I}$ is that of a round two-sphere of scalar curvature $k$.  In the
following, we will fix an allowed choice of $\Omega$ on $M \cup {\cal
I}$ and a choice of $k$. We will then take\footnote{Note that our
imposition of this rather rigid structure on $\cal F$ as a result of
our gauge fixing is not done merely for convenience, but is necessary
in order that ${\mbox{\boldmath $\omega$}}$ extend to $\cal I$.}
${\cal F}$ to consist of metrics, $g_{ab}$, on $M$ such that $\Omega^2
g_{ab}$ extends smoothly to ${\cal I}$ and equals $\tilde{g}^0_{ab}$
there, and such that the Bondi condition (\ref{bondi}) holds on ${\cal
I}$. It may then be checked that the general notion of infinitesimal
asymptotic symmetries given in the previous section corresponds to the
usual notion of infinitesimal BMS symmetries; indeed, our general
definition of infinitesimal asymptotic symmetries corresponds closely
to the definition of infinitesimal BMS symmetries\footnote{The only
difference between our definition and the definition given in
\cite{gw} concerns the notion of the equivalence of two
representatives, $\xi^a$ and ${\xi'}^a$. In addition to requiring
agreement of $\xi^a$ and ${\xi'}^a$ at $\cal I$, we impose the extra
requirement that they give rise to the same asymptotic integral
(\ref{I}). However, it is not difficult to show that if $\xi^a$ and
${\xi'}^a$ agree at $\cal I$ then they automatically give rise to the
same asymptotic integral (\ref{I}).} given in \cite{gw}.

It follows immediately from our conditions on ${\cal F}$ that the
unphysical perturbed metric
\begin{equation}
\tilde{\gamma}_{ab} \equiv \Omega^2 \gamma_{ab}
\label{gamtilde}
\end{equation}
extends smoothly to ${\cal I}$ and vanishes there, so it can be
written in the form
\begin{equation}
\tilde{\gamma}_{ab} = \Omega \tau_{ab}
\label{tau}
\end{equation}
where $\tau_{ab}$ extends smoothly to ${\cal I}$ and, in general, is
nonvanishing there. Furthermore, since $\delta n_a = 0$, we have
\begin{equation}
\delta [\tilde{\nabla}_a n_b] = - \{ \tilde{\nabla}_{(a}
\tilde{\gamma}_{b)c} - \frac{1}{2} \tilde{\nabla}_c
\tilde{\gamma}_{ab} \} n^c
\label{dbondi}
\end{equation}
Substituting from eqs.(\ref{gamtilde}),
(\ref{tau}), and (\ref{n}) and setting the resulting expression to
zero on ${\cal I}$ in accord with eq.(\ref{bondi}), we obtain
\begin{equation}
n_{(a} \tau_{b)c} n^c |_{\cal I} = 0
\label{taun}
\end{equation}
This, in turn, implies that $\tau_{bc} n^c$ vanishes on ${\cal I}$, so
we may write
\begin{equation}
\tau_{bc} n^c = \Omega \tau_b 
\label{taub}
\end{equation}
where $\tau_b$ is smooth (and, in general, nonvanishing) at ${\cal I}$.
This implies that
\begin{equation}
\delta n^a = \delta (\tilde{g}^{ab} n_b) = - \Omega \tau^{ab} n_b = -
\Omega^2 \tau^a
\label{dn}
\end{equation}

The crucial issue with regard to the applicability of the ideas of the
previous section is whether the presymplectic current
$3$-form\footnote{As noted in section 2, ${\mbox{\boldmath $\omega$}}$
has the ambiguity (\ref{Y2}). However, Iyer \cite{i} has shown
that if ${\bf Y}$ is such that ${\mbox{\boldmath $\theta$}}$ maintains
the general form given by eq.(23) of \cite{iw} with the coefficients
in that formula being regular, analytic functions of the fields, then
${\bf Y}$ must vanish on ${\cal I}$. Consequently, in vacuum general
relativity, the limit to ${\cal I}$ of ${\mbox{\boldmath $\omega$}}$
is, in fact, unique.} ${\mbox{\boldmath $\omega$}}$ extends
continuously to ${\cal I}$. To investigate this, we express the
quantities appearing in eq.(\ref{omegagr}) in terms of $\Omega$ and
variables that extend smoothly to ${\cal I}$. Clearly, the unphysical
volume element
\begin{equation}
\tilde {\mbox{\boldmath $\epsilon$}} = \Omega^4 {\mbox{\boldmath $\epsilon$}}
\label{tildeeps}
\end{equation}
and
\begin{equation}
\tilde{P}^{abcdef} \equiv \Omega^{-6} P^{abcdef}
\label{tildeP}
\end{equation}
extend smoothly to ${\cal I}$ and are nonvanishing there. We eliminate
the the action of the physical derivative operator, $\nabla_a$, on
$\gamma_{ab}$ in terms of the unphysical derivative operator,
$\tilde{\nabla}_a$, via
\begin{equation}
\nabla_a \gamma_{bc} = \tilde{\nabla}_a \gamma_{bc} + 2 {C^d}_{a(b}
\gamma_{c)d}
\label{der}
\end{equation}
where (see, e.g., \cite{w})
\begin{equation}
{C^c}_{ab} = 2 \Omega^{-1} {\delta^c}_{(a} n_{b)} - \Omega^{-1} n^c
\tilde{g}_{ab}
\label{C}
\end{equation}
Finally, we substitute 
\begin{equation}
\gamma_{ab} = \Omega^{-1} \tau_{ab} .
\label{gamtau}
\end{equation}
The terms appearing in the resulting expression for ${\mbox{\boldmath
$\omega$}}$ may now be classified as follows: (i) Terms in which
$\tilde{\nabla}_a$ acts on $\tau_{1ab}$ or $\tau_{2ab}$. For these
terms, the powers of $\Omega$ resulting from eqs.(\ref{tildeeps}),
(\ref{tildeP}), and (\ref{gamtau}) cancel, so these terms extend
smoothly to ${\cal I}$ and are, in general, nonvanishing there. (ii)
Terms in which $\tilde{\nabla}_a$ does not act on $\tau_{1ab}$ or
$\tau_{2ab}$ and $w^a$ is proportional to $n^a$. These terms cancel
due to the antisymmetry in $\tau_{1ab}$ and $\tau_{2ab}$. (iii) Terms
in which $\tilde{\nabla}_a$ does not act on $\tau_{1ab}$ or
$\tau_{2ab}$ but $w^a$ is not proportional to $n^a$. These terms
necessarily contain a contraction of $n^a$ with $\tau_{1ab}$ or
$\tau_{2ab}$, and eq.(\ref{taub}) can then be used. The extra power of
$\Omega$ picked up by the use of this equation ensures that these
terms extend smoothly to ${\cal I}$, where they are, in general,
nonvanishing. The upshot is that ${\mbox{\boldmath $\omega$}}$ extends
smoothly to ${\cal I}$ and is, in general, nonvanishing there. Thus,
with our definition of $\cal F$, asymptotically flat spacetimes at
null infinity in general relativity do indeed fall into the category
of ``case II'' of the previous section.

To apply the proposed prescription of the previous section to define a
``conserved quantity'', ${\cal H}_\xi$, for each BMS generator, $\xi^a$, and
each cross-section, $\partial \Sigma$, of ${\cal I}$, we need an
explicit formula for the pullback, $\bar{\mbox{\boldmath $\omega$}}$,
of the extension of ${\mbox{\boldmath $\omega$}}$ to ${\cal I}$. To do
so, we define ${}^{(3)} {\mbox{\boldmath $\epsilon$}}$ by
\begin{equation}
\tilde{\epsilon}_{abcd} = 4 \, {}^{(3)} \epsilon_{[abc} n_{d]}
\label{3eps}
\end{equation}
so that the pullback, ${}^{(3)} \bar{\mbox{\boldmath $\epsilon$}}$, of
${}^{(3)} {\mbox{\boldmath $\epsilon$}}$ to ${\cal I}$ defines a
positively oriented volume element\footnote{For past null infinity,
this volume element would be negatively oriented, resulting in sign
changes in some of the formulas below.} on ${\cal I}$ (see footnote
\ref{orient4}). We have
\begin{equation}
\bar{\mbox{\boldmath $\omega$}} = - \frac{1}{16 \pi} \Omega^{-4} n_a
w^a \, {}^{(3)} \bar{\mbox{\boldmath $\epsilon$}}
\label{omegabar}
\end{equation}
A lengthy but entirely straightforward calculation starting with
eq.(\ref{w}), making the substitutions (\ref{der})-(\ref{gamtau}), and
making heavy use of eqs.(\ref{bondi}), (\ref{n2}), and (\ref{taub})
yields (see also \cite{am}, \cite{abr})
\begin{equation}
\Omega^{-4} n_a w^a|_{\cal I} = \frac{1}{2} \{ - \tau_{2}^{bc} n^a
\tilde{\nabla}_a \tau_{1bc} + \tau_2 n^a \tilde{\nabla}_a \tau_1
+ \tau_2 n^a \tau_{1a} \} - [1 \leftrightarrow 2]
\label{omegascri}
\end{equation}
where we have written $\tau = {\tau^a}_a$ and ``$1 \leftrightarrow
2$'' denotes the same terms as in the preceding expression with $1$
and $2$ interchanged.

The above formula can be rewritten in a more useful form as
follows. By a direct computation using eq.(7.5.14) of \cite{w}, the
variation of the unphysical Ricci tensor at ${\cal I}$ is given by
\begin{equation}
\delta \tilde{R}_{ab} |_{\cal I} = - n_{(a} \tilde{\nabla}_{b)} \tau
- n^c \tilde{\nabla}_c \tau_{ab} + n_{(b} \tilde{\nabla}^d \tau_{a)d}
+ n_{(a} \tau_{b)}
\label{dR}
\end{equation}
Hence, defining $S_{ab}$ by
\begin{equation}
S_{ab} \equiv \tilde{R}_{ab} - \frac{1}{6} \tilde{R} \tilde{g}_{ab}
\label{S}
\end{equation}
we obtain
\begin{equation}
\delta S_{ab} |_{\cal I} =  - n_{(a} \tilde{\nabla}_{b)} \tau
- n^c \tilde{\nabla}_c \tau_{ab} + n_{(b}
\tilde{\nabla}^d \tau_{a)d} + n_{(a} \tau_{b)}
- \frac{1}{3} (-n^c \tilde{\nabla}_c \tau
+ n^c \tau_c) \tilde{g}_{ab}
\label{dS}
\end{equation}
On the other hand, $\tilde{R}_{ab}$ is related to $R_{ab}$ by the
usual conformal transformation formulae (see, e.g., Appendix D of
\cite{w}). Setting $R_{ab} = 0$ by the vacuum field equations, it
follows that (see eq.(6) of \cite{g})
\begin{equation}
S_{ab} = - 2 \Omega^{-1} \tilde{\nabla}_{(a} n_{b)} + \Omega^{-2} n^c
n_c \tilde{g}_{ab}
\label{Sconf}
\end{equation}
Taking the variation of this equation and evaluating the resulting
expression on $\cal I$ using eqs.(\ref{dbondi}), (\ref{tau}),
(\ref{dn}) and (\ref{taub}), we obtain
\begin{equation}
\delta S_{ab}|_{\cal I} = 4n_{(a} \tau_{b)} - n^c \tilde{\nabla}_c
\tau_{ab} - n^c \tau_c \tilde{g}_{ab}
\label{dSconf}
\end{equation}
Comparing this formula with eq.(\ref{dS}), we obtain
\begin{equation}
[\tilde{\nabla}^b \tau_{ab} - \tilde{\nabla}_a \tau - 3 \tau_a]
|_{\cal I} = 0
\label{ntau0}
\end{equation}
as well as
\begin{equation}
[n^b \tilde{\nabla}_b \tau + 2 n^b \tau_b] |_{\cal I} = 0
\label{ntau}
\end{equation}
Using eq.(\ref{ntau}) together with eq.(\ref{dS}), we see that
\begin{equation}
\Omega^{-4} n_a w^a|_{\cal I} = \frac{1}{2} [\tau_{2}^{ab} \delta_1
S_{ab} -\tau_{1}^{ab} \delta_2 S_{ab}]
\label{omegascri2}
\end{equation}
Now, the Bondi news tensor, $N_{ab}$, on ${\cal I}$ is defined by
\cite{g}
\begin{equation}
N_{ab} = \bar{S}_{ab} - \rho_{ab}
\label{news}
\end{equation}
where $\bar{S}_{ab}$ denotes the pullback to ${\cal I}$ of $S_{ab}$
and $\rho_{ab}$ is the tensor field on ${\cal I}$ defined in general
by eq.(33) of \cite{g}, which, in our gauge choice, is just
$\frac{1}{2} k \bar{g}^0_{ab}$, where $\bar{g}^0_{ab}$ denotes the
pullback to $\cal I$ of $\tilde{g}^0_{ab}$. Since $\delta \rho_{ab} =
0$ and since, by eq.(\ref{taub}), $\tau^{ab}$ on ${\cal I}$ is tangent
to ${\cal I}$, we may replace $\delta S_{ab}$ by $\delta N_{ab}$ in
eq.(\ref{omegascri2}). Thus, we obtain our desired final formula:
\begin{equation}
\bar{\mbox{\boldmath $\omega$}} = - \frac{1}{32 \pi} [\tau_{2}^{ab}
\delta_1 N_{ab} -\tau_{1}^{ab} \delta_2 N_{ab}] \, {}^{(3)}
\bar{\mbox{\boldmath $\epsilon$}}
\label{omegabar2}
\end{equation}

To apply our prescription, we must find a symplectic potential,
${\mbox{\boldmath $\Theta$}}$, for $\bar{\mbox{\boldmath $\omega$}}$
on ${\cal I}$ which is locally constructed\footnote{A major subtlety
would have arisen in the meaning of ``locally constructed'' if we had
not imposed the rigid background structure given by the Bondi
condition (\ref{bondi}) together with our fixing of
$\tilde{g}^0_{ab}$. If, say, the background structure was specified
merely by the ``asymptotic geometry'' as defined on P.22 of \cite{g},
then there would exist diffeomorphisms locally defined in the
neighborhood of a point $p \in {\cal I}$ which preserve the background
structure but cannot be extended to globally defined diffeomorphisms
which preserve the background structure. Indeed, a necessary condition
for a background-structure-preserving local diffeomorphism to be
globally extendible is that it preserve the tensor field $\rho_{ab}$,
defined by eq.(33) of \cite{g}, since $\rho_{ab}$ can be constructed
from a global specification of the background structure. Now, locally
defined diffeomorphisms that are not globally extendible are not
relevant to the definition of ``locally constructed'' given in
footnote \ref{loc}, since that definition requires globally defined
diffeomorphisms. Since the allowed (globally defined) diffeomorphisms
must locally preserve $\rho_{ab}$, that quantity would, in effect,
count as ``local'' with regard to the definition of ``local
construction'' of ${\mbox{\boldmath $\Theta$}}$---even though the
construction of $\rho_{ab}$ from the background structure given in
\cite{g} involves the global solution to differential
equations. Consequently, the Bondi news tensor (which is constructed
out of manifestly local quantities and $\rho_{ab}$) would still be
considered as ``locally constructed'' even if the background structure
had been specified as in \cite{g}. This subtlety does not arise here,
since with our gauge choice, $\rho_{ab}$ and the Bondi news tensor are
manifestly local.} out of the spacetime metric, $g_{ab}$, and
background structure (and depends analytically on the metric), is
independent of any arbitrary choices made in specifying the background
structure, and is such that ${\mbox{\boldmath $\Theta$}}(g_{ab},
\gamma_{ab})$ vanishes for all $\gamma_{ab}$ whenever $g_{ab}$ is
stationary. By inspection, a symplectic potential satisfying all of
these properties is given by\footnote{That $N_{ab}$ and hence
${\mbox{\boldmath $\Theta$}}$ vanish for all stationary solutions is
proven, e.g., on P.53-54 of \cite{g}.}
\begin{equation}
{\mbox{\boldmath $\Theta$}} = - \frac{1}{32 \pi} N_{ab} \tau^{ab} \,
{}^{(3)} \bar{\mbox{\boldmath $\epsilon$}}
\label{Thetascri}
\end{equation}

As discussed in section 3, this choice of ${\mbox{\boldmath
$\Theta$}}$ will be unique if and only if there does not exist a
$3$-form $\bf W$ on ${\cal I}$ which is locally constructed (in the
sense of footnote \ref{loc}) out of the physical metric, $g_{ab}$, and
background structure (and depends analytically on the physical
metric), is independent of any arbitrary choices made in specifying
the background structure, and is such that $\delta {\bf W}$ vanishes
for all $\gamma_{ab}$ whenever $g_{ab}$ is stationary. In our case,
the only relevant ``background structure'' present is the conformal
factor, $\Omega$, since all other background quantities (such as
$\tilde{g}^0_{ab}$ and $n^a$ on $\cal I$) can be reconstructed from
$\Omega$ and the physical metric.  Now, the physical metric, $g_{ab}$,
its curvature, ${R_{abc}}^d$, and (physical) derivatives of the
curvature all can be expressed in terms of the unphysical metric,
$\tilde{g}_{ab}$, its curvature, $\tilde{R}_{abc}{}^d$, and unphysical
derivatives of the unphysical curvature together with $\Omega$ and its
unphysical derivatives. Therefore, we may view $\bf W$ as a function
of $\tilde{g}_{ab}$, $\tilde{R}_{abc}{}^d$, and unphysical derivatives
of $\tilde{R}_{abc}{}^d$, together with $\Omega$ and its unphysical
derivatives. The requirement that $\bf W$ vary analytically with
$g_{ab}$ (at fixed $\Omega$) then implies that it must depend
analytically on $\tilde{g}_{ab}$ at fixed $\Omega$.

In our specification of conditions on the background structure, we
required that $\tilde{g}^0_{ab}$ induce a round two-sphere metric of
scalar curvature $k$ on all cross-sections of ${\cal I}$. The choice
of $k$ was arbitrary, and could have been fixed at any value.  If we
keep $\cal F$ fixed (i.e., consider the same class of physical
metrics) but change $\Omega$ by $\Omega \rightarrow \lambda \Omega$
with $\lambda$ constant, then $\tilde{g}^0_{ab}$ will induce a round
two-sphere metric of scalar curvature $\lambda^{-2} k$ rather than $k$
on all cross-sections. We require that under this scaling of $\Omega$
(corresponding to modifying an ``arbitrary choice'' in the
specification of the background structure), we have ${\bf W}
\rightarrow {\bf W}$.

To analyze the implications of this requirement, it is useful to
introduce the following notion of the {\it scaling dimension} \cite{g}
of a tensor, ${T^{a_1 ... a_k}}_{b_1 ... b_l}$, of type $(k,l)$ which
is locally constructed out of the unphysical metric and $\Omega$: If
under the scaling $\Omega \rightarrow \lambda \Omega$ keeping the
physical metric fixed we have ${T^{a_1 ... a_k}}_{b_1 ... b_l}
\rightarrow \lambda^p {T^{a_1 ... a_k}}_{b_1 ... b_l}$, then we define
the scaling dimension, $s$, of ${T^{a_1 ... a_k}}_{b_1 ... b_l}$ by
\begin{equation}
s = p + k - l
\label{scale}
\end{equation}
It follows that the scaling dimension of a tensor does not change under
the raising and lowering of indices using the unphysical metric. It is
easily seen that the scaling dimension of $\Omega$ is $+1$, the scaling
dimension of the unphysical metric is $0$, and the scaling dimension of the
unphysical curvature tensor is $-2$. Each derivative decreases the
scaling dimension by one, so, for example, the scaling dimension of $n_a =
\tilde{\nabla}_a \Omega$ is $0$ and the scaling dimension of the $j$th
derivative of the unphysical curvature is $-(j+2)$. 

Since the $3$-form $\bf W$ is required to be invariant under scaling
of $\Omega$, it must have a scaling dimension of $-3$. Since ${}^{(3)}
\epsilon_{abc}$ has scaling dimension $0$, if we define $w = W_{abc}
\, {}^{(3)} \epsilon^{abc}$ we obtain a scalar with scaling dimension
$-3$. By our assumptions, $w$ must be locally constructed out of
$\Omega$ and $\tilde{g}_{ab}$ (in the sense of footnote \ref{loc}) and
must vary analytically with $\tilde{g}_{ab}$ at fixed
$\Omega$. Presumably, this will imply that we can write $w$ as a
convergent sum of terms (with coefficients depending on the conformal
factor) of products (with all indices contracted) of the unphysical
metric, the unphysical curvature, unphysical derivatives of the
unphysical curvature, $n_a = \tilde{\nabla}_a \Omega$ and unphysical
derivatives of $n^a$. (Negative powers of $\Omega$ can, of course,
occur in the coefficients if they multiply a term which vanishes
suitably rapidly at $\cal I$.) Now, the unphysical metric, the
unphysical curvature, and $n_a$ all have have a non-positive scaling
dimension and derivatives only further decrease the scaling
dimension. Therefore, if any term were composed of more than two
factors containing the unphysical curvature tensor, the only way of
achieving a scaling dimension of $-3$ would be to multiply it by a
positive power of $\Omega$, in which case it would vanish at $\cal
I$. Similarly, if the term contained a single factor with two or more
derivatives of curvature, it also would have to vanish at $\cal
I$. Similar restrictions occur for terms containing derivatives of
$n^a$. This reduces the possible terms that can occur in $w$ to a
small handful, and it is then easily verified that there does not
exist an allowed $w$ such that $\delta w$ is nonzero in general (so
that it contributes nontrivially to ${\mbox{\boldmath $\Theta$}}$) but
$\delta w$ vanishes whenever the physical metric, $g_{ab}$, is
stationary. Therefore, we conclude that ${\mbox{\boldmath $\Theta$}}$
is unique.

To complete the prescription, we need to specify a stationary
``reference solution'' $\phi_0$ satisfying eq.(\ref{consis2}). A
natural candidate for $\phi_0$ is Minkowski spacetime and, indeed, it
should be possible to show that no other stationary
solution\footnote{If $t^a$ denotes the timelike Killing vector field,
then $\int_{\partial \Sigma} {\bf Q}[t]$ is proportional to the Komar
formula for mass and is nonvanishing for all stationary solutions
other than Minkowski spacetime. We expect that eq.(\ref{consis2}) will
fail when $\eta^a$ is an asymptotic boost and $\xi^a$ is an asymptotic
spatial translation such that their commutator yields $t^a$.} can
satisfy eq.(\ref{consis2}). In Minkowski spacetime, an arbitrary
infinitesimal asymptotic symmetry can be written as a sum of a Killing
vector field plus a supertranslation. Since eq.(\ref{consis2}) holds
automatically whenever either $\eta^a$ or $\xi^a$ is a Killing vector
field, it suffices to check eq.(\ref{consis2}) for the case where both
$\eta^a$ and $\xi^a$ are supertranslations, i.e., on $\cal I$ they are
of the form $\xi^a = \alpha n^a$, $\eta^a = \beta n^a$ where $\alpha$
and $\beta$ are such that $n^a \tilde{\nabla}_a \alpha = n^a
\tilde{\nabla}_a \beta = 0$. Since the satisfaction of
eq.(\ref{consis2}) does not depend upon the choice of representative
of the infinitesimal asymptotic symmetry, we may assume that $\eta^a$
and $\xi^a$ satisfy the Geroch-Winicour gauge condition \cite{gw}
$\nabla_a \eta^a = \nabla_a \xi^a = 0$ (see below). In that case,
$\int_{\partial \Sigma} {\bf Q}[{\cal L}_\eta \xi]$ will vanish and
eq.(\ref{consis2}) reduces to
\begin{equation}
0 = \int_{\partial \Sigma} \{ \eta \cdot {\mbox{\boldmath
$\theta$}}(\phi_0,{\cal L}_\xi \phi_0) - \xi \cdot {\mbox{\boldmath
$\theta$}}(\phi_0,{\cal L}_\eta \phi_0) \}
\label{consis3}
\end{equation}
From eq.(\ref{thetagr}) we obtain on $\cal I$
\begin{equation}
\eta^c \theta_{cab}(\phi, \delta \phi) = \frac{1}{16 \pi}
\tilde{\epsilon}_{abcd} V^c \eta^d
\label{thetab}
\end{equation}
where
\begin{equation}
V^a \equiv \Omega^{-1} [\tilde{\nabla}_b \tau^{ab} - \tilde{\nabla}^a
\tau - 3 \tau^a]
\label{V2}
\end{equation}
and it should be noted that $V^a$ has a smooth limit to $\cal I$ on
account of eq.(\ref{ntau0}). The pullback of $\eta \cdot
{\mbox{\boldmath $\theta$}}$ to $\cal I$ is thus
\begin{equation}
\eta \cdot \bar{\mbox{\boldmath $\theta$}} = - \frac{1}{16 \pi} \beta
n_a V^a n \cdot {}^{(3)} \bar{\mbox{\boldmath $\epsilon$}}
\label{ptheta}
\end{equation}
In using this equation to evaluate the term $\eta \cdot
{\mbox{\boldmath $\theta$}}(\phi_0,{\cal L}_\xi \phi_0)$ in
eq.(\ref{consis3}), we must substitute $\chi_{ab}$ for $\tau_{ab}$
where
\begin{eqnarray}
\chi_{ab} & \equiv & \Omega {\cal L}_\xi g_{ab} \nonumber \\
& = & \Omega^{-1} [{\cal L}_\xi \tilde{g}_{ab} - 2K \tilde{g}_{ab}]
\label{chi}
\end{eqnarray}
with
\begin{equation}
K \equiv \Omega^{-1} \xi^a n_a
\label{K}
\end{equation}
(Thus, $\chi_{ab} = 2X_{ab}$ in the notation of \cite{gw}; it follows
directly from the definition of infinitesimal asymptotic symmetries
that $\chi_{ab}$ and $K$ extend smoothly to $\cal I$.) It may then be
seen by inspection of eq.(19) of \cite{gw} that ${\mbox{\boldmath
$\theta$}}(\phi_0,{\cal L}_\xi \phi_0)$ is proportional to the
``linkage flux'' (see below) associated with $\xi^a$. However, from
the formula for the linkage flux for supertranslations in Minkowski
spacetime given in eq.(10) of \cite{aw}, it may be verified that that
$\int_{\partial \Sigma} \eta \cdot {\mbox{\boldmath
$\theta$}}(\phi_0,{\cal L}_\xi \phi_0)$ cancels $\int_{\partial
\Sigma} \xi \cdot {\mbox{\boldmath $\theta$}}(\phi_0,{\cal L}_\eta
\phi_0)$, so eq.(\ref{consis3}) is indeed satisfied, as we desired to
show.

Thus, for the case of null infinity in general relativity, the general
prescription proposed in section 4 instructs us to define a
``conserved quantity'', ${\cal H}_\xi$, for each infinitesimal BMS
symmetry $\xi^a$ and each cross-section, $\partial \Sigma$, of $\cal I$
by
\begin{equation}
\delta {\cal H}_\xi = \int_{\partial \Sigma} [\delta {\bf Q} - \xi
\cdot {\mbox{\boldmath $\theta$}}] - \frac{1}{32 \pi} \int_{\partial
\Sigma} N_{ab} \tau^{ab} \xi \cdot {}^{(3)} \bar{\mbox{\boldmath
$\epsilon$}}
\label{dhfinal}
\end{equation}
together with the requirement that ${\cal H}_\xi = 0$ for all $\xi^a$
and all cross-sections in Minkowski spacetime. 

By our above arguments, there exists a unique ${\cal H}_\xi$
satisfying the above requirements. How does this prescription compare
with the one previously given by Dray and Streubel \cite{ds}? From our
general analysis of section 4, it follows that our prescription
automatically yields the flux formula
\begin{equation}
{\bf F}_\xi = {\mbox{\boldmath $\Theta$}} (g_{ab}, {\cal L}_\xi
g_{ab}) = - \frac{1}{32 \pi} N_{ab} \chi^{ab} \, {}^{(3)}
\bar{\mbox{\boldmath $\epsilon$}}
\label{ffinal}
\end{equation}
Equation
(\ref{ffinal}) agrees with the flux formula proposed by Ashtekar and
Streubel \cite{as} (see eq.(19) of \cite{aw}). But it was shown by
Shaw \cite{s} that the Dray-Streubel prescription also yields the
Ashtekar-Streubel flux formula. Therefore, the difference between our
${\cal H}_\xi$ and the ``conserved quantity'' proposed by Dray and
Streubel must be a quantity that depends locally on the fields at the
cross-section $\partial \Sigma$ and yet---since the flux associated
with the difference of these quantities vanishes---for a given
solution, is independent of the choice of cross-section (i.e., this
difference, if nonzero, would be a truly conserved quantity). If we restrict
attention to spacetimes that are asymptotically flat at both null and
spatial infinity, the equivalence of our prescription to that of Dray
and Streubel would follow from the fact that they both yield the ADM
conserved quantities in the limit as the cross-section approaches
spatial infinity. However, it is instructive to show the
equivalence of the two prescriptions directly (without assuming
asymptotic flatness at spatial infinity), and we now turn our
attention to doing so.

Let $\partial \Sigma$ be a cross-section of ${\cal I}$ and let $\xi^a$
be a representative of an infinitesimal asymptotic symmetry (i.e., an
infinitesimal BMS representative). We may uniquely decompose $\xi^a$
into a part that is everywhere tangent to $\partial \Sigma$ on
$\partial \Sigma$ plus a supertranslation. Since both our prescription
and that of Dray and Streubel are linear in $\xi^a$, it suffices to
consider the equivalence of the prescription for each piece
separately, i.e., to consider separately the cases where (a) $\xi^a$
is everywhere tangent to $\partial \Sigma$ and (b) $\xi^a$ is a
supertranslation.

Consider, first, case (a), where as discussed in section 4, a true
Hamiltonian exists. In case (a), eq.(\ref{dhfinal}) is simply
\begin{equation}
\delta {\cal H}_\xi = \int_{\partial \Sigma} \delta {\bf Q}
\label{dhfinala}
\end{equation}
One might think that the solution to this equation would be simply
${\cal H}_\xi = \int_{\partial \Sigma} {\bf Q}$, which corresponds to
the Komar formula with the correct numerical factor for angular
momentum (see eq.(\ref{Qgr}) above). However, although $\int_{\partial
\Sigma} \delta {\bf Q}$ is well defined and independent of choice of
infinitesimal BMS representative $\xi^a$ (as it must be according to
the general considerations of section 4), it was shown in \cite{gw}
that the value of $\int_{\partial \Sigma} {\bf Q}$ depends upon the
choice of infinitesimal BMS representative, and, in this sense, is ill
defined unless a representative is specified. It was also shown in
\cite{gw} that the Geroch-Winicour condition $\nabla_a \xi^a = 0$ in
$M$ (where $\nabla_a$ is the {\it physical} derivative operator) picks
out a class of representatives which makes $\int_{\partial \Sigma}
{\bf Q}$ well defined. (By eq.(\ref{chi}), the Geroch-Winicour
condition is equivalent to $\chi = 0$, where $\chi = \tilde{g}^{ab}
\chi_{ab}$.) We write ${\bf Q}_{GW}$ to denote ${\bf Q}$ when $\xi^a$
has been chosen so as to satisfy the Geroch-Winicour condition. It was
shown in \cite{gw} that $\int_{\partial \Sigma} {\bf Q}_{GW}$ is
equivalent to a previously proposed ``linkage formula'' \cite{tw} for
defining ``conserved quantities''. Furthermore, this linkage formula
has the property that when $\xi^a$ is everywhere tangent to $\partial
\Sigma$, it yields zero in Minkowski spacetime\footnote{This fact
follows immediately from the equivalence of eqs.(21) and (22) of
\cite{s}.} as desired. This suggests that the solution to
eq.(\ref{dhfinala}) together with the requirement that ${\cal H}_\xi$
vanish in Minkowski spacetime is ${\cal H}_\xi = \int_{\partial
\Sigma} {\bf Q}_{GW}$.  However, it is far from obvious that this
formula satisfies eq.(\ref{dhfinala}), since when we vary the metric,
we also must, in general, vary $\xi^a$ in order to continue to satisfy
the Geroch-Winicour gauge condition, $\chi = 0$. Indeed, under a
variation of the metric, $\delta \tilde{g}_{ab} = \Omega \tau_{ab}$,
keeping $\xi^a$ fixed it follows from eq.(\ref{chi}) that
\begin{eqnarray}
\delta \chi & = & \delta (\tilde{g}^{ab}
\chi_{ab}) = \delta(\Omega^{-1} g^{ab} {\cal L}_\xi g_{ab})
\nonumber \\
& = & \Omega^{-1} {\cal L}_\xi \gamma = \Omega^{-1} {\cal L}_\xi (\Omega \tau)
\nonumber \\
& = & {\cal L}_\xi \tau + K \tau
\label{dchi}
\end{eqnarray}
where, as previously defined above, $\tau = \tilde{g}^{ab}
\tau_{ab}$. Consequently, in order to preserve the Geroch-Winicour
condition, it will be necessary to vary the infintesimal BMS
representative by $\delta \xi^a = \Omega^2 u^a$ (see \cite{gw}) where
$u^a$ satisfies
\begin{equation}
2 \Omega^{-1} \nabla_a (\Omega^2 u^a) = - {\cal L}_\xi \tau - K \tau
\label{u}
\end{equation}
Since $\nabla_a u^a = \tilde{\nabla}_a u^a - 4 \Omega^{-1} u^a n_a$,
this relation can be expressed in terms of unphysical variables as
\begin{equation}
2 \Omega \tilde{\nabla_a} u^a - 4 u^a n_a = - {\cal L}_\xi
\tau - K \tau
\label{u2}
\end{equation}

Clearly, we have
\begin{equation}
\delta \int_{\partial \Sigma} {\bf Q}_{GW} = \int_{\partial \Sigma}
\delta {\bf Q} - \frac{1}{16 \pi} \int_{\partial \Sigma}
\epsilon_{abcd} \nabla^c (\Omega^2 u^d)
\label{dQGW}
\end{equation}
where $u^a$ satisfies eq.(\ref{u2}). We wish to show that the second
term on the right side of eq.(\ref{dQGW}) vanishes. To do so, it is
convenient to introduce a null vector field $l^a$ as follows. At
points of $\partial \Sigma$ we take $l^a$ to be the unique
(past-directed) null vector that is orthogonal to $\partial \Sigma$
and satisfies $l^a n_a = 1$. We extend $l^a$ to all of $\cal I$ by
requiring that ${\cal L}_n l^a = 0$ on $\cal I$. Finally, we extend
$l^a$ off of $\cal I$ via the geodesic equation $l^b \tilde{\nabla}_b
l^a = 0$. A calculation similar to that given in eq.(17) of \cite{gw}
shows that the integrand of the second term in eq.(\ref{dQGW}) can be
written as
\begin{eqnarray}
I_{ab} & \equiv & \epsilon_{abcd} \nabla^c (\Omega^2 u^d) \nonumber \\
& = & [l^c \tilde{\nabla}_c Y + \frac{1}{2} Y \tilde{\nabla}_c l^c + \tilde{D}_c s^c] \,\, {}^{(2)} \tilde{\epsilon}_{ab}
\label{Iab}
\end{eqnarray}
where $s^a$ denotes the projection of $u^a$ to $\partial \Sigma$;
$\tilde{D}_a$ and ${}^{(2)} \tilde{\epsilon}_{ab}$ are the derivative
operator and volume element on $\partial \Sigma$ associated with the
induced unphysical metric, $\tilde{q}_{ab}$, on $\partial \Sigma$; and
we have written
\begin{equation}
Y \equiv  \frac{1}{2} [{\cal L}_\xi \tau + K \tau]
\label{Y}
\end{equation}
The term $\tilde{D}_c s^c$ is a total divergence and integrates to
zero\footnote{It is erroneously stated in \cite{gw} that $\tilde{q}_{ab}
\tilde{\nabla}^a u^b$ is an intrinsic divergence. The dropping of that
term does not affect any of the results in the body of that
paper. However, the formula given in footnote 20 of \cite{gw} is valid
only when $\chi$ ($=2X$ in the notation of \cite{gw}) vanishes on
$\cal I$.}. After a significant amount of algebra, it can be shown
that the remaining terms in eq.(\ref{Iab}) can be expressed as
\begin{equation}
{\bf I}' = \frac{1}{2} {\cal L}_\xi [({\cal L}_l \tau
+\frac{1}{2} \tau \tilde{\nabla}_a l^a) \,\, {}^{(2)}
\tilde{\mbox{\boldmath $\epsilon$}}]
\label{Iab2}
\end{equation}
These remaining terms integrate to zero since $\xi^a$ is tangent to
$\partial \Sigma$. This establishes that
\begin{equation}
\delta \int_{\partial \Sigma} {\bf Q}_{GW} = \int_{\partial \Sigma}
\delta {\bf Q}
\label{QQGW}
\end{equation}
and thus the unique solution to eq.(\ref{dhfinala}) which vanishes in
Minkowski spacetime is
\begin{equation}
{\cal H}_\xi = \int_{\partial \Sigma} {\bf Q}_{GW}
\label{Ha}
\end{equation}
which is equivalent to the linkage formula. This agrees with the
Dray-Streubel expression in case (a).

We turn our attention now to case (b) where $\xi^a$ is a
supertranslation and thus takes the form \cite{gw}
\begin{equation}
\xi^a = \alpha n^a - \Omega \tilde{\nabla}^a \alpha + O(\Omega^2)
\label{st}
\end{equation}
where $\alpha$ is such that on $\cal I$ we have $n^a \tilde{\nabla}_a
\alpha = 0$. Direct substitution of (\ref{st}) into the variation of
eq.(\ref{Qgr}) yields on $\cal I$ \cite{i}
\begin{equation}
\delta Q_{ab} = - \frac{1}{16 \pi} \tilde{\epsilon}_{abcd}
\tilde{\nabla}^c (\alpha \tau^d - \tau^{de} \tilde{\nabla}_e \alpha) 
\label{dQb}
\end{equation}
from which it follows that the pullback, $\delta \bar{\bf Q}$, of
$\delta {\bf Q}$ to $\cal I$ is given by
\begin{equation}
\delta \bar{\bf Q} = - \frac{1}{16 \pi} U \cdot
{}^{(3)} \bar{\mbox{\boldmath $\epsilon$}}
\label{pdQb}
\end{equation}
where
\begin{equation}
U^a = \tilde{\nabla}^a (\alpha \tau^b n_b) - \alpha n^b
\tilde{\nabla}_b \tau^a - n^a \tau^b \tilde{\nabla}_b \alpha +
n_b \tilde{\nabla}_c \alpha \tilde{\nabla}^b \tau^{ac}
\label{U}
\end{equation}
The pullback of $\xi \cdot {\mbox{\boldmath $\theta$}}$ to $\cal I$ is
given by eq.(\ref{ptheta}) above (with the substitutions $\eta
\rightarrow \xi$ and $\beta \rightarrow \alpha$).

Thus, our general prescription instructs us to define ${\cal H}_\xi$
in case (b) by the requirement that ${\cal H}_\xi = 0$ in Minkowski
spacetime together with the equation
\begin{equation}
\delta {\cal H}_\xi = - \frac{1}{16 \pi} \int_{\partial \Sigma} [U^a
l_a - \alpha V^a n_a + \frac{1}{2} \alpha N_{ab}
\tau^{ab}] \,\, {}^{(2)} {\mbox{\boldmath $\epsilon$}}
\label{dhfinalb}
\end{equation}
where $l_a$ is any covector field on $\cal I$ satisfying $n^a l_a =
1$.  A lengthy calculation \cite{i} shows that the solution
to this equation is the expression given by Geroch \cite{g}, namely
\begin{equation}
{\cal H}_\xi = \frac{1}{8 \pi} \int_{\partial \Sigma} P^a
l_a  \,\, {}^{(2)} {\mbox{\boldmath $\epsilon$}}
\label{hfinalb}
\end{equation}
where
\begin{equation}
P^a = \frac{1}{4} \alpha K^{ab} l_b + (\alpha {D}_b l_c + l_b
{D}_c \alpha) \bar{g}^{cd} N_{de} \bar{g}^{e[b} n^{a]}
\label{Pg}
\end{equation}
Here ${D}_a$ is the derivative operator on $\cal I$ defined on P.46-47
of \cite{g}; $\bar{g}^{ab}$ is the (non-unique) tensor field on $\cal
I$ satisfying $\bar{g}_{ac} \bar{g}^{cd} \bar{g}_{db} = \bar{g}_{ab}$
where $\bar{g}_{ab}$ denotes the pullback to $\cal I$ of
$\tilde{g}_{ab}$; and $K^{ab} = {}^{(3)} \bar{\epsilon}^{acd} \,
{}^{(3)} \bar{\epsilon}^{bef} \Omega^{-1} \bar{C}_{cdef}$ where
$\Omega^{-1} \bar{C}_{cdef}$ denotes the pullback to $\cal I$ of the
limit to $\cal I$ of $\Omega^{-1} \tilde{C}_{cdef}$, where
$\tilde{C}_{cdef}$ denotes the unphysical Weyl tensor. Equation
(\ref{hfinalb}) agrees with the Dray-Streubel prescription in case
(b). Consequently, our prescription agrees with that given by Dray and
Streubel for all infitesimal BMS representatives $\xi^a$ and all
cross-sections $\partial \Sigma$, as we desired to show.

\section{Summary and Outlook}

In this paper, using ideas arising from the Hamiltonian formulation,
we have proposed a general prescription for defining notions of
``conserved quantities'' at asymptotic boundaries in diffeomorphism
covariant theories of gravity. The main requirement for the
applicability of our ideas is that the symplectic current $(n-1)$-form
${\mbox{\boldmath $\omega$}}$ extend continuously to the boundary. If,
in addition, the pullback of ${\mbox{\boldmath $\omega$}}$ vanishes at
the boundary (``Case I''), then a Hamiltonian associated with each
infinitesimal asymptotic symmetry exists, and the value of the
Hamiltonian defines a truly conserved quantity. On the other hand, if
the pullback of ${\mbox{\boldmath $\omega$}}$ fails to vanish in
general at the boundary (``case II''), our prescription requires us to
find a symplectic potential on the boundary which vanishes for
stationary solutions. When such a symplectic potential exists and is
unique---and when a ``reference solution'' $\phi_0$ can be found
satisfying the consistency condition (\ref{consis})---we have provided
a well defined prescription for defining a ``conserved quantity'',
${\cal H}_\xi$, for each infinitesimal asymptotic symmetry, $\xi^a$ and
cross-section $\partial \Sigma$. This ``conserved quantity'' is
automatically local in the fields in an arbitrarily small neighborhood
of the cross-section and has a locally defined flux given by the
simple formula (\ref{F5}). For the case of asymptotically flat
spacetimes at null infinity in vacuum general relativity, our proposal
was shown to yield a unique prescription which, furthermore, was shown
to agree with the one previously given by Dray and Streubel \cite{ds}
based upon entirely different considerations. In this way, we have
provided a link between the Dray-Streubel formula and ideas arising
from the Hamiltonian formulation of general relativity.

Since our approach does not depend on the details of the field
equations---other than that they be derivable from a diffeomorphism
covariant Lagrangian---there are many possible generalizations of the
results we obtained for vacuum general relativity. We now mention some
of these generalizations, all of which are currently under
investigation.

Perhaps the most obvious generalization is to consider asymptotically
flat spacetimes at null infinity in general relativity with matter
fields, $\psi$, also present. If the asymptotic conditions on $\psi$
are such that the ${\mbox{\boldmath $\omega$}}$ continues to extend
continuously to $\cal I$ and are such that the physical stress-energy
tensor, $T_{ab}$, satisfies the property that $\Omega^{-2} T_{ab}$
extends continuously to $\cal I$ (so that ``$T_{ab}$ vanishes
asymptotically to order $4$'' in the terminology of \cite{g}), then an
analysis can be carried in close parallel with that given in section 5
for the vacuum case.  For minimally coupled fields (i.e., fields such
that the curvature does not explicitly enter the matter terms in the
Lagrangian), it follows from the general analysis of \cite{iw} that
there will be no matter contributions to $\bf Q$ from the term ${\bf
X}^{ab} \nabla_{[a} \xi_{b]}$ (see eq.(\ref{Qform}) above). (Even for
non-minimally coupled fields such as the conformally invariant scalar
field, the ${\bf X}^{ab} \nabla_{[a} \xi_{b]}$ term in $\bf Q$ will
retain the vacuum form (\ref{Qgr}) in the limit as one approaches
$\cal I$.)  However, in general the symplectic potential
${\mbox{\boldmath $\theta$}}$ and symplectic current ${\mbox{\boldmath
$\omega$}}$ will pick up additional contributions due to the matter
fields and the other terms in $\bf Q$ in eq.(\ref{Qform}) may also
acquire matter contributions. For the massless Klein-Gordon scalar
field, $\psi$, we require $\Omega^{-1} \psi$ to have a smooth limit to
$\cal I$. In that case, ${\mbox{\boldmath $\omega$}}$ extends
continuously to $\cal I$. Although $T_{ab}$ does not actually vanish
asymptotically to order $4$ in this case (see the Appendix of
\cite{qw}), it appears that all the essential features of the analysis
of section 5 carry through nonetheless. In Einstein-Klein-Gordon
theory no additional matter terms occur in $\bf Q$, so $\bf Q$
continues to be given by eq.(\ref{Qgr}). Furthermore, the extension to
$\cal I$ of the pullback to surfaces of constant $\Omega$ of the
matter field contribution to ${\mbox{\boldmath $\theta$}}$ satisfies
the property that it vanishes for stationary solutions. Consequently,
in this case we can define ${\mbox{\boldmath $\Theta$}}$ on $\cal I$
by simply adding this additional matter contribution to
${\mbox{\boldmath $\theta$}}$ to the right side of
eq.(\ref{Thetascri}). The upshot is that the explicit matter
contributions to formula (\ref{dhfinal}) cancel, so that ${\cal
H}_\xi$ is again given by the linkage formula (\ref{Ha}) when $\xi^a$
is tangent to $\partial \Sigma$ and is given by eq.(\ref{hfinalb})
when $\xi^a$ is a supertranslation. However, the flux formula
(\ref{ffinal}) will pick up additional terms arising from the
additional matter contributions to ${\mbox{\boldmath $\theta$}}$ and
hence to ${\mbox{\boldmath $\Theta$}}$. Similar results hold for
non-minimally-coupled scalar fields, such as the conformally coupled
scalar field\footnote{For Maxwell and Yang-Mills fields, a new issue
of principle arises as a result of the additional gauge structure of
these theories. If we merely require the vector potential $A_a$ to
extend smoothly to $\cal I$, then ${\mbox{\boldmath $\omega$}}$ will
extend continuously to $\cal I$ and, by the general analysis of
section 4, the integral defining $\Omega_\Sigma$ will always
exist. However, $\Omega_\Sigma$ will not be gauge invariant. (Thus, a
Hamiltonian on $\Sigma$ conjugate to gauge transformations will fail
to exist in general in much the same way as a Hamiltonian conjugate to
infinitesimal asymptotic symmetries fails to exist in general.)
Consequently, in these cases it appears that substantial gauge fixing
at $\cal I$ would be needed in order to obtain gauge invariant
expressions for ``conserved quantities''.}.

The analysis is similar in the case of higher derivative gravity
theories if we {\it impose}, in addition to the usual asymptotic
conditions at null infinity, the requirement that $\Omega^{-2} R_{ab}$
extends continuously to $\cal I$. (Of course, there is no guarantee
that the field equations will admit a reasonable number of solutions
satisfying this property.) If we consider a Lagrangian which, in
addition to the Einstein-Hilbert term (\ref{Lgr}), contains terms
which are quadratic and/or higher order in the curvature and its
derivatives, then additional terms will appear in ${\bf Q}$ as well as
${\mbox{\boldmath $\theta$}}$ and ${\mbox{\boldmath $\omega$}}$ (see
\cite{iw}). However, it appears that none of these additional terms
will contribute to ${\cal H}_\xi$ or its flux when the limit to $\cal
I$ is taken. Thus, it appears that the formulas for both the
``conserved quantities'' and their fluxes will be the same in higher
derivative gravity theories as in vacuum general
relativity\footnote{The interpretation of this result would be that,
although higher derivative gravity theories may have additional
degrees of freedom, these extra degrees of freedom are massive and do
not propagate to null infinity (and/or they give rise to instabilities
and are excluded by our asymptotic assumptions).}.

Our proposal also can be applied to situations where the asymptotic
conditions considered are very different from those arising in vacuum
general relativity. Thus, for example, it should be possible to use
our approach to define notions of total energy and radiated energy in
dilaton gravity theories in $2$-dimensional spacetimes. It should also
be possible to use our approach for asymptotically anti-deSitter
spacetimes in general relativity with a negative cosmological
constant. When suitable asymptotic conditions are imposed, the
asymptotically anti-deSitter spacetimes should lie within ``Case I''
of section 4, so it should be possible to define truly conserved
quantities conjugate to all infinitesimal asymptotic symmetries. It
would be of interest to compare the results that would be obtained by
our approach with those of previous approaches \cite{ads}.

Finally, we note that many of the ideas and constructions of section 4
would remain applicable if $\cal B$ were an ordinary timelike or null
surface $\cal S$ in the spacetime, $M$, rather than an asymptotic
boundary of $M$. Thus, one could attempt to use the ideas presented
here to define notions of quasi-local energy contained within $\cal S$
and/or energy radiated through $\cal S$. However, it seems unlikely
that a unique, natural choice of ${\mbox{\boldmath $\Theta$}}$ will
exist in this context, so it seems unlikely that this approach would
lead to a unique, natural notion of quasi-local energy. Nevertheless,
by considering the case where $\cal S$ is the event horizon of a black
hole, it is possible that the ideas presented in this paper may
contain clues as to how to define the entropy of a nonstationary black
hole in an arbitrary theory of gravity obtained from a diffeomorphism
covariant Lagrangian.

\section{Acknowledgements}

This research was initiated several years ago by one of us (R.M.W.)
and Vivek Iyer. Some unpublished notes provided to us by Vivek Iyer
\cite{i} (dating from an early phase of this research) were extremely
useful in our investigations. Some unpublished calculations by Marc
Pelath for a scalar field in Minkowski spacetime were useful for
refining our proposal. We have benefitted from numerous discussions
and comments from colleagues, particularly Abhay Ashtekar and Robert
Geroch. We wish to thank Abhay Ashtekar, Piotr Chrusciel, and Roh Tung
for reading the manuscript. This research was supported in part by NSF
grant PHY 95-14726 to the University of Chicago and by a NATO
scholarship from the Greek Ministry of National Economy.

\end{document}